\documentclass[structabstract]{aa}  
\usepackage{graphicx}
\usepackage{amstext} 
\usepackage{amssymb} 
\usepackage{natbib} 
\bibpunct{(}{)}{;}{a}{}{,} 
\usepackage{txfonts}
\usepackage[urlcolor=blue,colorlinks=true,linkcolor=black,citecolor=black]{hyperref}
\usepackage{breakurl}
\begin{document}
   \title{The PRIMA fringe sensor unit\thanks{\textit{Part of this work is based on technical observations collected at the European Southern Observatory at Paranal, Chile. Public data can be downloaded at \url{http://www.eso.org/sci/activities/vltcomm/prima/PACMAN_CommDataRelease_text.html}.}}}
\author{J. Sahlmann
		\inst{1,2}
		\and
		S. M\'enardi
		\inst{1}
		\and R. Abuter
		\inst{1}
		\and M. Accardo
		\inst{1}
		\and S. Mottini
		\inst{3}
		\and
		F. Delplancke
		\inst{1}}		
\institute{European Southern Observatory, Karl-Schwarzschildstrasse 2, 85748 Garching, Germany\\		
		\email{johannes.sahlmann@unige.ch}	
	\and
		Observatoire de Gen\`eve, 51 Chemin Des Maillettes, 1290 Sauverny, Switzerland
	\and
	      Thales Alenia Space Italia, Strada Antica di Collegno 253, 10146 Turin, Italy}
\date{Received 3 April 2009 / Accepted 3 September 2009} 
\abstract
{The fringe sensor unit (FSU) is the central element of the phase referenced imaging and micro-arcsecond astrometry (PRIMA) dual-feed facility and provides fringe sensing for all observation modes, comprising off-axis fringe tracking, phase referenced imaging, and high-accuracy narrow-angle astrometry. It is installed at the Very Large Telescope Interferometer (VLTI) and successfully served the fringe-tracking loop during the initial commissioning phase.}
{To maximise sensitivity, speed, and robustness, the FSU is designed to operate in the infrared {\it K}-band and to include spatial filtering after beam combination and a very-low-resolution spectrometer without photometric channels. It consists of two identical fringe sensors for dual-star operation in PRIMA astrometric mode.}
{Unique among interferometric beam combiners, the FSU uses spatial phase modulation in bulk optics to retrieve real-time estimates of fringe phase after spatial filtering. The beam combination design accommodates a laser metrology for pathlength monitoring. An $R=20$ spectrometer across the {\it K}-band makes the retrieval of the group delay signal possible. The calibration procedure uses the artificial light source of the VLTI laboratory and is based on Fourier transform spectroscopy to remove instrumental effects.}
{The FSU was integrated and aligned at the VLTI in July and August 2008. It yields phase and group delay measurements at sampling rates up to 2~kHz, which are used to drive the fringe-tracking control loop. During the first commissioning runs, the FSU was used to track the fringes of stars with {\it K}-band magnitudes as faint as $m_K=9.0$, using two VLTI auxiliary telescopes (AT) and baselines of up to 96~m. Fringe tracking using two Very Large Telescope (VLT) unit telescopes was demonstrated.} 
{The concept of spatial phase-modulation for fringe sensing and tracking in stellar interferometry is demonstrated for the first time with the FSU. During initial commissioning and combining stellar light with two ATs, the FSU showed its ability to improve the VLTI sensitivity in {\it K}-band by more than one magnitude towards fainter objects, which is fundamental for achieving the scientific objectives of PRIMA.}
\maketitle
\section{Introduction}
Fringe tracking is essential for increasing the efficiency and accuracy of astronomical observations with optical interferometers. Because it is  the equivalent of adaptive optics for single-dish telescopes, fringe tracking partly removes the atmospheric piston turbulence that limits the quality of interferometric measurements, at the cost of decreased sensitivity. After the first demonstration by \cite{Shao1980}, fringe tracking has become a common feature of modern optical interferometers and opened the path to new scientific results \citep{Monnier2003, Barry2008, Bouquin2009a}. It makes narrow-angle astrometry of the order of $100~\mu$as possible, which is demonstrated both in single-beam interferometry \citep{Colavita1994, Lane:2004rm} and later in a dual-beam interferometer, where the technique is extended to off-axis fringe tracking and phase referencing \citep{Lane:2003zl}. A new instrument relying on fringe-tracking observations is PRIMA \citep{Quirrenbach:1998mi, Delplancke2006}.\\ 
The deployment of the PRIMA system at the VLTI \citep{Haguenauer2008} began in July 2008. Once fully operational, the PRIMA dual-beam facility will increase the sensitivity of the VLTI and enable it to perform phase-referenced imaging and high-accuracy narrow-angle astrometry. During PRIMA operation, two objects within an isoplanatic angle are simultaneously observed with two unit telescopes (UT) or auxiliary telescopes (AT), and their light is routed in four separated beams by star separator modules \citep{Nijenhuis:2008cy} at the telescopes.\\ 
In astrometric mode, the two beam pairs are combined in the twin fringe sensors of the FSU: FSUA and FSUB, each combining two beams originating from the primary object and the secondary object, respectively. In normal mode, FSUB serves as fringe tracker on the brighter, primary object whereas FSUA observes the fringes of the fainter, secondary object. In swap mode, introduced to eliminate systematic and instrumental effects, it is the opposite. An infrared laser metrology system \citep{Schuhler2007} measures the internal differential optical path difference (OPD) between the two observed objects and a set of differential delay lines (DDL, \citealt{Pepe2008}) equalises the differential sidereal delay between the two objects. Eventually, the angular separation between both objects can be derived from OPD measurements in FSUA and FSUB and the differential OPD delivered by the laser metrology \citep{Delplancke2006, Elias:2008sf}.\\
In imaging mode, PRIMA provides two-telescope off-axis fringe tracking and the phase reference for the present VLTI-instruments \citep{Quirrenbach:1998mi}. There, the light of the primary object is fed into one of the FSU fringe sensors, which drives the fringe-tracking control loop. The light of the secondary object, potentially much dimmer than the primary object, is combined in either AMBER \citep{Petrov2007} or MIDI \citep{Leinert2003}.\\    
In all modes, the FSU plays the central role within the PRIMA facility. Two identical FSU fringe sensors can deliver real-time estimates of phase, group delay, and signal-to-noise ratio (SNR) for one or two observed targets. Hence, it serves both as the scientific instrument for astrometry and as sensor for off-axis fringe tracking. The requirements imposed by these operation modes lead to the high-throughput, high-bandwidth twin design of the FSU.\\
The current magnitude limit for fringe tracking with FINITO at the VLTI in routine operation with the ATs is $m_H=5$ \citep{ESO2009} and slightly fainter in engineering mode ($m_H=6$, \citealt{Bouquin2008}). Using the FSU, it is possible to extend this limit to a magnitude of $m_K \sim 7 \-- 8$. We describe the instrument concept and implementation, report on laboratory test results, and present first on-sky performances.
\begin{figure*}
\includegraphics[width= \linewidth]{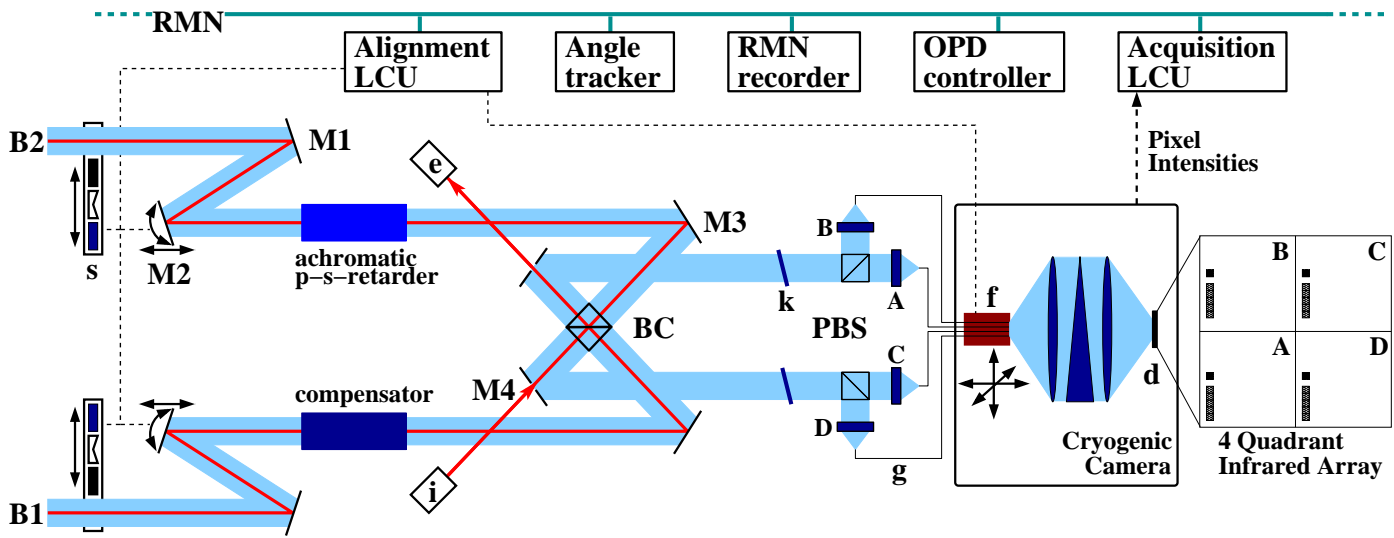}
\caption{FSU layout showing the optical beam path, the beam combination and detection principle, remotely controlled optics, and the control system structure: two telescope beams (B1 and B2) enter the FSU from the VLTI laboratory and motorised stages (s) are used to shut the beams or to introduce longitudinal atmospheric dispersion compensators. The mirrors M1, M3, and M4 are fixed, whereas M2 is a piezoelectrically controlled tip-tilt mirror and mounted on a motorised translation stage for longitudinal pathlength adjustment. Both beams have been superimposed in the beam combiner (BC) after one beam experienced the achromatic retardation of $\pi / 2$ between p- and s-polarisation. Dichroic mirrors (k) deflect the \textit{H}-band light. A polarising beam splitter (PBS) splits each combined beam and the four generated beams (A, B, C, and D) separated by $\pi / 2$ in phase, respectively, are injected in single-mode fibres (g) by means of coupling doublets. The cold optics image the fibre bundle (f) on the four-quadrant infrared detector (d) after spectral dispersion. The fibre bundle is mounted onto a 2-axis linear piezoelectric stage for lateral image alignment, and its focus position is adjustable with a manually controlled stepper motor. Raw pixel intensities are delivered to the acquisition local control unit (LCU). The shutter stages, the M2 actuators, and the fibre bundle lateral position are piloted by the alignment LCU. Laser metrology beams (red) are injected (i) and extracted (e) in the beam combiner, pass through a central dichroic patch on M4 and propagate in the centre of the telescope beams. The FSU LCUs communicate with the other VLTI real-time control systems via the reflective memory network (RMN). The layout is identical for FSUA and FSUB.}
\label{fig:fsu_principle}
\end{figure*}
\section{Scientific Rationale}
The astrophysical science accessible with PRIMA can be summarised with respect to the operation mode. One of the scientific drivers of the astrometric mode is the extrasolar planet search and characterisation, where a large programme is being prepared \citep{Launhardt2008}. This programme exploits the potential of narrow-angle astrometry with dual-beam interferometers \citep{Shao1992} and the expected accuracy of  $30-40~\mu$as with PRIMA \citep{Belle2008}. Combining astrometric with spectroscopic, radial-velocity observations of planetary systems can yield the complete orbital solution, hence solve for the planets' mass \citep{Benedict:2002rz, Bean2007, Pravdo:2009rz}. Similarly, astrometric observations can yield the components' masses of binary systems \citep{Muterspaugh:2005lq, Muterspaugh:2008yg, Lane:2007sf} and can be used to explore the dynamics of the stellar cluster close to the galactic centre's black hole \citep{Bartko2008}.\\
In the off-axis fringe-tracking and imaging mode, PRIMA can be used to explore stellar surfaces and circumstellar discs, as well as to carry out high-resolution observations of active galactic nuclei, making use of the visibility measurement capabilities of AMBER and MIDI, together with the phase reference provided by PRIMA and the improvement in limiting magnitude of this mode \citep{Quirrenbach:1998mi}.   
\begin{figure*}
\includegraphics[width= \textwidth]{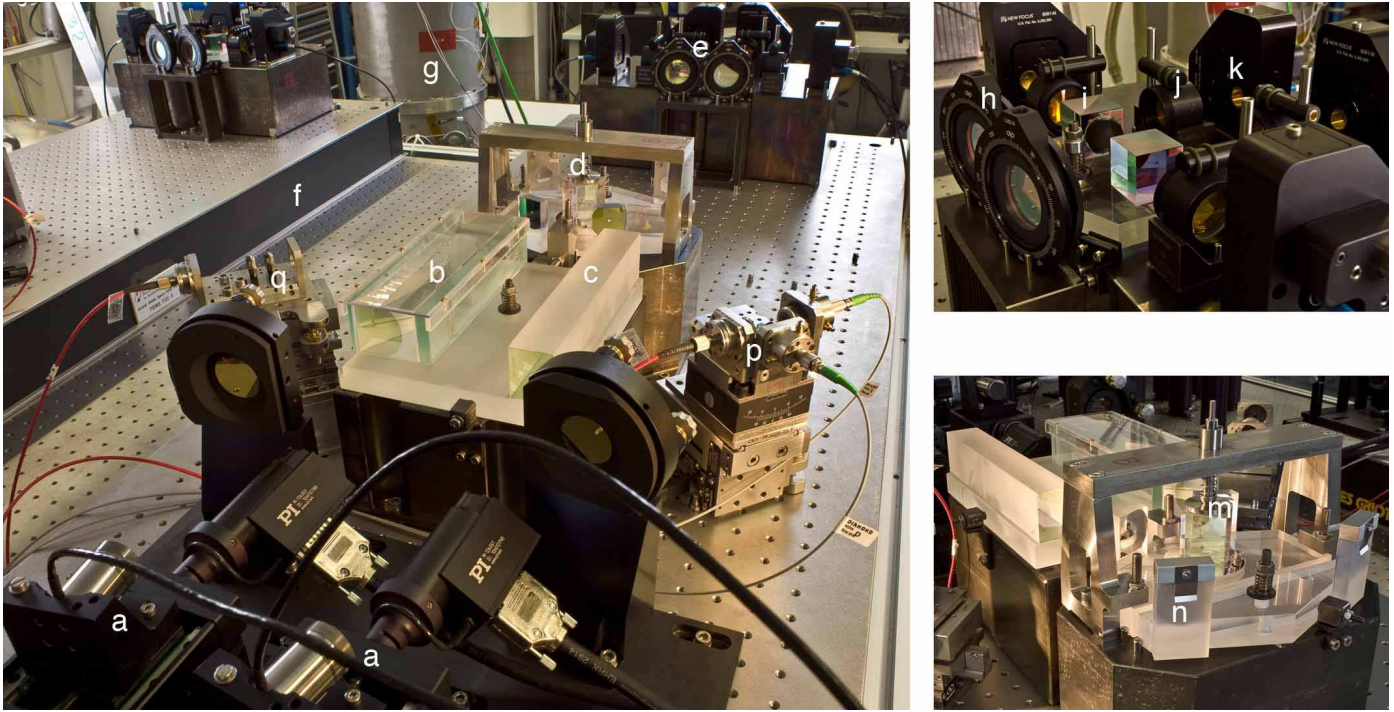}
\caption{FSUB breadboard with opto-mechanics installed in the testbed: {\it Left}: alignment system with tip-tilt piezo and linear motor (a), achromatic retarder (b), optical path length compensator (c), beam combiner (d), and spatial filter system (e). The FSUA breadboard (f) and the cryostat (g) are partially visible. Laser metrology injection (p) and extraction (q) optics are installed on the FSU breadboard. The shutter system is not shown. {\it Top right}: Close-up of the spatial filters showing {\it H}/{\it K}-dichroics (h),  polarising beam-splitters (i), coupling doublets (j), and fibre positioners (k). {\it Bottom right}: Close-up of the beam combining system with beam combiner (m) and folding mirrors (n) (Photos: ESO/H.-H.~Heyer).} \label{fig:breadboard}
\end{figure*}
\section{FSU specifications}\label{sec:spec}
The FSU design is driven by three key criteria: 
\begin{enumerate}
\item High throughput to reach faint stellar magnitudes,
\item High bandwidth and robustness against OPD and wavefront perturbations for correction of atmospheric and instrumental piston disturbances via fringe tracking, 
\item Twin design of FSUA and FSUB to allow PRIMA to swap objects to eliminate systematic errors in astrometric mode.
\end{enumerate}
These lead to the following specifications for the FSU.
\begin{itemize}
	\item Fringe sensing is performed in the atmospheric {\it K}-band ($1.95 \, \mu$m - $2.45\, \mu$m). The effective wavelength is longer than in {\it H}-band, hence the non-ambiguous phase measurement range of this bandpass is 35 \% wider. In addition, the visibility loss due to longitudinal atmospheric dispersion is two times less than in  {\it H}-band, and the expected Strehl ratio, hence fibre-coupling efficiency is higher. 
	\item Spatial phase modulation is applied in contrast to a temporal phase modulation scheme. The delay estimates are delivered at the sampling rate of the detector. 
	\item The design does not include photometric channels to accommodate a common combiner for telescope beams and for the laser metrology and to maximise the optical throughput, hence increase the sensitivity. Common telescope and metrology beam combination is a requirement set by the astrometric mode.
	\item Operation is possible at a selectable sampling rate between 0.5~Hz and 2~kHz. 
	\item FSUA and FSUB are opto-mechanically and electronically identical systems and their roles are interchangeable for the accurate calibration of astrometric observations.
\end{itemize}	 
\section{Instrument description}
The FSU was initially designed and manufactured by Thales Alenia Space Italy in cooperation with Osservatorio Astronomico di Torino \citep{Mottini2005, Gai2004} following the technical specifications requested by the European Southern Observatory (ESO). It was delivered to ESO in July 2006 where it was installed in a dedicated test laboratory, referred to as the testbed \citep{Abuter2006}. Consequent to the findings during the testing period, important changes to FSU hardware and software had to be carried out until the delivery of the instrument to the VLTI observatory in July 2008 \citep{Sahlmann2008a}.  
\subsection{Phase measurement}
The common technique for obtaining the phase measurements of interfering light is to sample the fringe at two or more points with known separation in phase space. The fringe phase relative to a reference can then be retrieved with standard formulae \citep{Wyant1975, Creath1988}.\\
In stellar interferometry there are currently two schemes to achieve the sampling of the fringe phase: spatial and temporal phase modulation. In the case of temporal modulation, the fringe packet is scanned at a fast rate ($\sim$100 Hz), typically using a piezo-driven actuator. Synchronising the detector read-out and the modulator produces a temporal sequence of detector reads across the fringe and the phase can be retrieved. Currently operating fringe trackers, such as FATCAT at the Keck interferometer \citep{Vasisht2003}, FINITO at VLTI \citep{Bouquin2008}, their equivalents at Palomar testbed interferometer \citep{Colavita1999a}, and the CHARA interferometer \citep{Berger2008}, apply temporal phase modulation. In the presence of fast piston perturbations, e.g., caused by vibrations in the beam relaying optics, this scheme suffers from the scrambling of the fringes during the acquisition sequence. Because the OPD seen by the sensor changes at frequencies comparable to the modulation frequency, the phase separation of consecutive reads is unknown, so that the phase measurement is of poor quality.\\
The FSU is the first fringe-tracking sensor to implement spatial phase modulation at an astronomical interferometer. Here the phase is modulated by static optical components, and four combined beams with known phase separation, denoted A, B, C, and D are produced. The four beam intensities are measured contemporaneously, and the fringe phase can be estimated after each detector read-out. Piston perturbations at frequencies comparable to the read-out rate ($\sim$1 kHz), which typically have very low amplitudes, reduce the fringe contrast, hence increase the noise, but do not distort the phase measurement.\\
We can can compare both schemes in two scenarios.
\begin{itemize}
\item The integration time per ABCD bin is identical. Spatial modulation has a 4 times greater measurement bandwidth than temporal modulation; however, the phase noise is 4 times higher for spatial modulation because the pixel intensity is reduced by a factor of 4. 
\item The phase sampling time is identical. The theoretical bandwidth is identical for both schemes, but the temporal scheme requires a 4 times faster detector and at least the reading of 2 pixels, corresponding to the two interferometric outputs, with the effect of slightly higher phase noise (factor $\sqrt{2}$).
\end{itemize}
The general drawback of the spatial scheme is that it has to account for the differential effect between the four channels, such as differential injection into the single-mode fibres, non-ideal phase separations, differential transmission, effective wavelengths, and pixel response. This calls for an accurate calibration of these instrumental effects.
\subsection{FSU opto-mechanical design}     
The FSU opto-mechanics are installed on a steel honeycomb optical table in the VLTI interferometric laboratory. The FSU footprint is a 1550$\times$1740~mm rectangle hosting two 600$\times$1500~mm breadboards, a cylindrical cryostat with $\sim$350~mm diameter and $\sim$600~mm height, and the shutter system, which is placed in front of the breadboards. FSUA and FSUB are identical in design and their respective warm optics are each mounted on one breadboard, whereas they share the cryostat for their cold optics and detectors (Fig. \ref{fig:breadboard}). Therefore, the description below applies to both of them.\\
Telescope beams are combined in bulk optics, where the spatial phase modulation is introduced. Four beams with a relative phase separation of $\pi / 2$ provide ABCD signals, used to reconstruct phase and group delay. Single-mode fibres spatially filter the beams and route the light to the cryogenic infrared camera, which includes a low-resolution spectrograph to obtain fringes in one white-light pixel, covering the full {\it K}-band, and five spectral pixels in each ABCD channel. The phase $\phi$ is computed from the white-light pixels, while the group delay is estimated using the spectral pixels.\\
Most optical components rely on reference pins for positioning to minimise manual alignment operations. Wherever possible, movable optics are motorised and controlled remotely with the result of 13 manual adjustment points, which are the four X-Y-Z fibre positioners and the focus position of the cold camera.\\
The FSU opto-mechanical system can be divided into three main components: the dispersion compensator and alignment system, the beam combination and spatial filter system, and the cryostat with spectrograph and detector.
\subsubsection{Dispersion compensator} 
Motorised stages are used to insert optical components in the input beams before beam combination (Fig. \ref{fig:fsu_principle}). Four configurations are available for each beam: 
\begin{description}
\item[S0] The beam path is free.
\item[S1] A black metal plate stops the beam and the detector sees an uniform background.
\item[S2] A retro-reflector oriented towards the beam combiner interrupts the beam: the detector sees itself and the laser metrology beam is retro-reflected.
\item[$\mathrm{S3}_t$] The beam passes through an infrared silica plate of thickness $t$ to
compensate for longitudinal atmospheric dispersion (LAD).
\end{description}
The glass plates $\mathrm{S3}_t$ are introduced to minimise the effects of LAD, mainly on fringe visibility \citep{Tango:1990rm, Leveque:1996rz}, when the total delay in air of the interferometer is longer than 24~m. Four thicknesses $t$ are available, two for each beam, and one unit of $t$ equals 1.7~mm, which is the thickness required to compensate for the LAD of 48~m beam-path in air (for typical VLTI environmental conditions, cf. \citealt{Daigne:1999sf}). The sign of the total delay is defined by the difference of optical path length in beam B1 and beam B2. The effective LAD compensator thickness $\Delta t$ is given by the thickness difference of $\mathrm{S3}_t$ in B1 and B2. Table \ref{table:ladc} summarises the LAD compensator configurations. With this setup offering five configurations, the LAD compensation is not dynamic and the required configuration has to be defined beforehand with the constraint that it cannot be changed during an observation.\\ 
LAD compensation in the FSU was not tested on the sky, and all results presented here were obtained without LADC. The effect of the inserted glass plates on the laser metrology beams was found not to be problematic in the testbed.
\begin{table} 
\caption{Configurations for LAD compensation} 
\label{table:ladc} 
\centering  
\begin{tabular}{r@{--}r c c r } 
\hline\hline 
\multicolumn{2}{c}{Total delay} &  B1 & B2  & $\Delta t$ \\
\multicolumn{2}{c}{(m)} & &  &   \\ 
\hline 
$-120\:$ & $\:-72$ & $\mathrm{S3}_4$ & $\mathrm{S3}_2$ & +2  \\ 
$  -72\:$ & $\:-24$  & $\mathrm{S3}_4$ & $\mathrm{S3}_3$ & +1  \\ 
$  -24\:$ & $\:+24$ & S0 & S0 & 0  \\ 
$ +24\:$ & $\:+72$  & $\mathrm{S3}_1$ & $\mathrm{S3}_2$ & -1  \\ 
$ +72\:$ & $\:+120$ & $\mathrm{S3}_1$ & $\mathrm{S3}_3$ & -2  \\ 
\hline  
\end{tabular} 
\end{table} 
\subsubsection{Beam alignment system}
The alignment system for each telescope beam consists of one static flat mirror (M1 in Fig. \ref{fig:fsu_principle}) and a second mirror (M2) mounted onto a piezo-driven tip-tilt stage (\textit{PI} S-330), which is attached to a motorised linear stage (\textit{PI} M-126) for OPD adjustment. This system was adapted from FINITO \citep{Bouquin2008} and makes fast (actuator bandwidth $>$~100~Hz) beam tip-tilt control possible. This functionality is essential for fringe-tracking operation at the VLTI, because it is required for automated injection optimisation and feed-forward correction of real-time tip-tilt measurements \citep{Bonnet2006, Sahlmann2007a}. The linear stage is needed to adjust the OPD when observing the calibration source and is driven by a DC motor, which makes it unsuitable as a fringe tracking actuator. \\
The shutter stages, the linear motors, and the piezoelectric stages of M2 are controlled by the alignment LCU (cf. Fig. \ref{fig:fsu_principle} and Sect. \ref{sec:lcu}). Consequently, each input beam can be remotely aligned with the FSU in terms of tip-tilt and OPD. The lateral pupil position during PRIMA observations is maintained by an independent control system, which is based on the PRIMA laser metrology and actuators in the star-separator modules \citep{Schuhler2007}.
\subsubsection{Achromatic phase shifter}\label{sec:ac}
Before the telescope beams are combined, an achromatic phase shifter introduces a $\pi / 2$ phase shift between the p- and s-polarisation component of one beam, achieved through three internal total reflections in a K-shaped silica prism \citep{Mottini2005}. The other beam traverses a silica block, acting as compensator to equalise the optical path in silica of both beams.\\
The optical path length of 207.6~mm inside the phase shifter and the compensator match within $50~\mu$m. The total optical path length difference inside silica is inferior to $100~\mu$m between both beams due to manufacturing errors of the phase shifter, compensator, and beam-combiner. As a result, the corresponding chromatic dispersion can account for at most a 0.75~\% contrast loss over the \textit{K}-band.\\   
Silica dispersion also accounts for $1.3~\mu$m differential OPD between $2.25~\mu$m and the laser metrology wavelength at $1.319~\mu$m. During PRIMA astrometric observations, this constant offset can be calibrated if the temperature gradient between phase shifter and compensator is small. To keep the offset variation below 1~nm over 30~min, the gradient has to remain constant within $0.04$~K, which is ensured by the VLTI laboratory thermal stability and by the good thermal connection of the phase shifter and compensator, which are glued on a common glass plate. 
\subsubsection{Beam combiner and spatial filter system}\label{sec:bc}
Both telescope beams are superimposed in the beam combiner, which is a 50/50 beam splitter cube, to produce two combined beams. Each combined beam is relayed to a polarising beam splitter, which separates the orthogonal p- and s-polarisations, and four combined beams are produced (Fig. \ref{fig:fsu_principle}). The {\it H}/{\it K}-dichroics, initially included with the main purpose of feeding a possible {\it H}-band FSU\footnote{This idea was abandoned for cost reasons after the design phase of the project.}, have a transmission of 0.05 \% at $\lambda = 1319$~nm and serve as filter to remove metrology laser straylight. Although they also attenuate by 1~\% in \textit{K}-band, they are kept in the optical path to limit the amount of background photons, especially for integration times of seconds, as envisioned on the secondary object in PRIMA astrometric mode. All transmissive bulk optics in the warm beamtrain are made of infrared silica. The combination of phase shifts originating from the achromatic phase shifter and from reflection or transmission in the beam combiner results in a relative phase shift of $\pi / 2$ between adjacent beams (Fig. \ref{fig:fringes}). These are the ABCD channels required for the phase computation. However, the measured phase shifts of the real system can depart from the ideal values by up to $\pi / 4$. Each of those ABCD beams is focussed by a coupling doublet onto the core of a single-mode fibre, which is held by a manual fibre positioner (Fig. \ref{fig:breadboard}). The coupling losses due to static aberrations caused by the doublets amount to ($9\pm1$)\% \citep{Sahlmann2007a}.\\
The beam propagation distance from the entrance of the FSU at the shutter to beam combination is $\sim$1300~mm and $\sim$2100~mm until the injection in the fibres. The fluoride glass fibres manufactured by \textit{Le Verre Fluor\'e} have a core diameter of 8.5 $\mu$m with a cut-off wavelength of 1.75~$\mu$m and a length of 3.25~m. The four fibres pass a vacuum feedthrough and guide the light of the ABCD channels into the cryostat and towards the cold optics.\\
According to the specifications in Sect. \ref{sec:spec}, the FSU beam combiner is also the injection point for the PRIMA laser metrology, explained in Sect. \ref{sec:calibmet}. During observation, the telescope pupils are imaged on the FSU beam combiner, independently of the variable propagation distance in the VLTI optical train. This is achieved with a variable curvature mirror located in the VLTI delay line cat's eye \citep{Ferrari:2003qe} and, for PRIMA observations, in the star separator modules \citep{Nijenhuis:2008cy}. Dynamical pupil re-imaging is a feature of VLTI \citep{Haguenauer2008} and is required to avoid beam vignetting by the FSU beam combiner, in spite of its minimum physical size. For PRIMA observations, it ensures that the laser metrology beam, returning from the star separator module at the telescope, remains focussed at the level of the FSU beam combiner, to minimise laser straylight on the detector.\\
After installation in the VLTI laboratory, long-term drift measurements of the FSU warm opto-mechanical parts were performed by regularly monitoring of the relative positions of the four warm fibre ends. The relative fibre positions can be deduced from the individual point spread function profiles, which are obtained by injecting one calibration source beam, modulating the M2 tilt mirror and recording the corresponding injected flux in each fibre along with the mirror tilt. Over 30 days, the maximum relative drift between two fibres of FSUA was measured to be smaller than 1~\% of the theoretical AT point spread function FWHM ($240\,\arcsec$ on sky in \textit{K}-band), which corresponds to approximately two times the alignment accuracy reached. No systematic drift was detected.
\subsubsection{Cryogenic low resolution spectrograph}\label{sec:lrs}
\begin{figure}
\includegraphics[width= \linewidth]{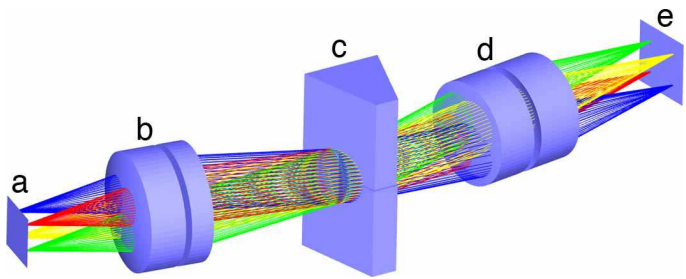}
\caption{Ray tracing through the FSU cold optics: fibre outputs (a), collimation doublet (b), prism assembly (c), camera lens doublet (d), and images on detector (e). Dispersed and undispersed images cannot be distinguished in this illustration. The propagation distance from (a) to (e) is 110~mm.} \label{fig:coldRays}
\end{figure}

\begin{figure}
\includegraphics[width= 6cm]{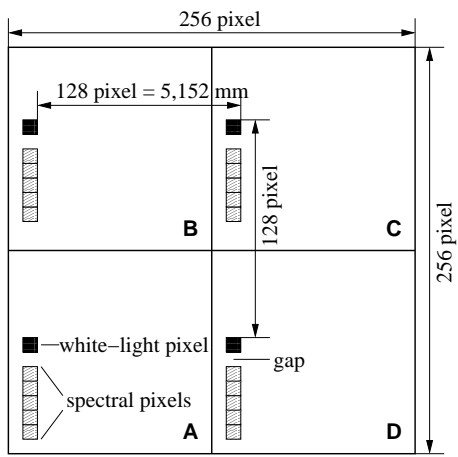}
\caption{Sketch of the PICNIC detector layout for the FSU (not to scale). The white-light pixels (filled) and the spectral pixels (striped) separated by a one pixel wide gap are indicated for each detector quadrant A,B,C, and D. Each pixel measures 40$\times$40$\,\mu$m. }
\label{fig:picnic}
\end{figure}

The outputs of the four fibres are assembled in a bundle to form a square array and are imaged on the detector by the cold optics, consisting of three optical components (Fig. \ref{fig:coldRays}). After collimation, the extended beam traverses the prism assembly, where two non-cemented prisms imposing different dispersion are mounted in contact. The front face of both prisms is located close to a pupil plane, and each prism intercepts a fraction of the pupil area. This fraction, defining the flux splitting ratio between the spectral pixels and the white-light pixel, can be manually adjusted by moving the common prism mount along the Y-axis. Both prisms apply nearly the same $16.3^\circ$ deviation to the incident collimated beams, but one prism, made of infrared silica, has $\sim$5 times higher dispersion than the other prism, made of barium fluoride ($\mathrm{BaF}_2$). A camera doublet then focusses the light on the four-quadrant infrared detector. Because of the different prism dispersions, one part of the beam is chromatically dispersed over the 5 spectral pixels, whereas the other part of the beam is dispersed over less than one pixel, providing the white-light pixel. Spectral and white-light pixels are aligned in a row and separated by a one pixel wide gap (Fig. \ref{fig:picnic}).\\
Collimating doublet and camera doublet lenses are made of infrared silica and zinc selenide (ZnSe). The silica prism carries a filter coating reflecting wavelengths above $2.45 \, \mu$m to reduce the thermal background, while the $\mathrm{BaF}_2$ prism has an anti-reflection coating. Both prisms are also used as filtering elements and have a notch filter coating that rejects the 1319~nm wavelength of the metrology laser. 
\subsection{Cryogenic system, actuators and detector}
\begin{figure}
\includegraphics[width= \linewidth]{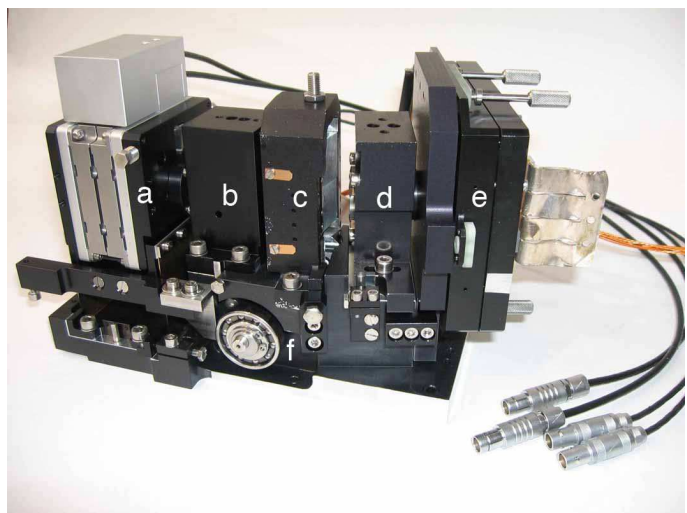}
\caption{FSU cold optics assembly: piezo X-Y-stage (a), collimator housing (b), prism assembly (c), camera lens housing (d), detector housing (e), and focus motor (f). The optical axis is horizontal in this image and light is propagating from left to right.}
\label{fig:coldOptics}
\end{figure}

The FSU cryostat contains the liquid nitrogen tank in the upper part and the cold optical bench in the lower part.  
FSUA and FSUB share a single cryostat; i.e., the cryostat hosts two identical sets of cold optics and two detectors. In operational condition, the cryostat is evacuated and cooled with liquid nitrogen. The fibre cables coming from the FSU warm optics pass a feedthrough and the bundle with the four fibre outputs is mounted on a piezo-driven X-Y translation stage (\textit{Piezosystem Jena}) with a range of 140~$\mu$m, used to finely align the fibre images on the pixels. The housings for collimation lenses, the prism assembly, the camera lens, and the detector are mounted on a common baseplate to  assure mechanical stability (Fig. \ref{fig:coldOptics}). Focus adjustment along the optical Z-axis is realised with a \textit{Phytron} stepper motor driving a spring-loaded flexure blade mechanism, which holds the piezo stage.\\
In the prism assembly, the two prisms are stacked vertically and held by an aluminium frame. The flux ratio between the white-light pixel and the spectral pixels is adjusted by sliding this frame vertically within the external housing. This operation requires the opening of the cryostat.\\
Each FSUA and FSUB uses one \textit{Teledyne} $256 \times 256$ PICNIC detector with four simultaneously sampled $128 \times 128$ pixel quadrants and a pixel size of 40$\times$40~$\mu$m. Each quadrant is used for one ABCD-channel. Only a subset of six pixels (five spectral pixels and one white-light pixel) in one column per quadrant is read, which reduces the read-out time compared to the full frame read-out by a factor of $10^{3}$. A total number of 24 detector pixels are therefore used for each fringe sensor (Fig. \ref{fig:picnic}). They are located close to the quadrant's origin to minimise the access time of pixel reads. The detector is read using a non-destructive read-out mode, where the electric charge accumulated in each pixel is evenly sampled $N$ times between two consecutive resets and the final pixel intensity is computed from a linear fit to the $N$ subsamples. For a typical FSU integration time of 1~ms, $N$ is limited to 16 by the detector video signal bandwidth. This leads to a read-out noise of $(19 \pm 1)$ electrons RMS in the testbed \citep{Sahlmann2008a}.\\
In addition to the flux ratio adjustment possibility, white-light and spectral pixels can be read at different rates, which provides high-bandwidth phase measurements and low-bandwidth group delay estimates with reduced noise. The white-light pixel can be read at a high rate for accurate phase tracking, whereas the spectral pixels can be read at a lower rate and the resulting group delay signal is used for fringe centring and the fringe acquisition. However, this mode was never tested because, during commissioning, the flux ratio had to be adjusted to put all light into the spectral pixels (see Sect. \ref{sec:firstfringes}).\\
The group delay estimate is extremely sensitive to relative drifts of the fibre bundle with respect to the detector in the direction of dispersion, because they alter the spectral content of the pixels. To achieve the envisioned astrometric precision of PRIMA, their effect on the measurement has to be negligible on a typical observation timescale of 30~minutes. An astrometric measurement is based on the difference of group delay measured by FSUA and FSUB, while observing reference and target object, hence it is sensitive to the instrumental drifts. The image motion was measured in the testbed to be below 2~$\mu$m over 40~hours for both axes and is compliant with the stability requirement over 30~minutes.
\subsection{Acquisition and control electronics}\label{sec:lcu}
The FSU PICNIC detector read-out is managed with the IRACE system \citep{Meyer1998} developed by ESO. FSU real-time computation and control at kHz data rates is performed by local control units (LCUs) based on \textit{Motorola} mv2700 CPU boards running VxWorks operating system, whereas high-level coordination and bookkeeping is done with Linux workstations. FSUA and FSUB run identical software and each comprise two real-time computers: one for acquisition and one for driving the alignment system. All LCUs use the reflective memory network (RMN) for fast communication and to interface the VLTI control system (Fig. \ref{fig:fsu_principle}), e.g. the angle tracker, the OPD controller, and the RMN recorder \citep{Abuter2008}. LCU time synchronisation at $\mu$s-level relies on the ESO fibre-optics timing system.\\ 
The acquisition LCU collects the non-destructive pixel reads to compute the pixel intensities and implements the real-time algorithms described in Sect. \ref{sec:algo}. The alignment LCU controls the actuators of cold and warm optics. Each LCU is housed in a VME crate, together with the interface cards to IRACE and the RMN. Optical links are used for transmission of the detector trigger signal and for piezo actuator control.
\subsection{Calibration source and laser metrology}\label{sec:calibmet}
Although they are not part of the FSU, the PRIMA laser metrology and the VLTI calibration source play an important role in the laboratory calibration (Sect. \ref{seq:labcal}).\\
Together with PRIMA, a new calibration and alignment unit was integrated in the VLTI laboratory. It generates four near-infrared beams, each covering the \textit{J}-, \textit{H}-, and \textit{K}-bands, using wavefront division (in contrast to the former unit which used amplitude division), providing a better control of spectrum, spatial coherence, and polarisation state \citep{Morel:2004si}. This unit was previously used in the testbed \citep{Sahlmann2007a} and is usually powered by a black body light source.\\
The role of the metrology system \citep{Schuhler2007} during PRIMA observations is to provide the internal differential OPD between the two observed objects. It uses super-heterodyne laser interferometry at $\lambda = 1319$~nm to measure the differential OPD between FSUA and FSUB $\Delta L_{A-B}$ and the internal OPD of FSUB $\Delta L_{B}$ in an incremental way. The metrology beams have a diameter of 1~mm at injection and propagate within the central obscuration of the telescope beams, which originates from the telescope secondary mirror. Metrology endpoints are given by the FSU beam combiners and the star separator modules \citep{Nijenhuis:2008cy}. Two laser beams, one for FSUA and FSUB respectively, with frequency-shifted orthogonal polarisation states are injected in and extracted from the stellar beams via centro-circular dichroic patches with 2.5~mm diameter on the FSU M4 mirrors (Figs. \ref{fig:fsu_principle} and \ref{fig:breadboard}). They propagate from the beam-combiners to the telescopes, where they are retro-reflected by the star separator modules,  and the return beams are extracted in the beam combiners. The metrology hardware includes the laser source, a custom frequency-stabilisation system \citep{Schuhler2007}, the injection and extraction optics, and three real-time computers interfacing the reflective memory network. The metrology estimates of $\Delta L_{A-B}$ and $\Delta L_{B}$ can therefore be recorded with the RMN recorder, but they are not used for feedback control.\\   
During the FSU laboratory calibration, the laser metrology monitors the OPD scan applied to the beams coming from the calibration unit and feeding FSUA and FSUB. For this purpose, the calibration unit provides metrology endpoints by means of integrated corner-cube retro-reflectors of 4~mm diameter, located in the central part of the beam behind a holed mirror and hence only acting on the metrology beams.
\subsection{Real time algorithms}\label{sec:algo}
The FSU real-time algorithms for phase, group delay, and SNR are evaluated based on the 24 instantaneous pixel intensities $I_{i,\Gamma}(t)$ at time $t$, where $i \in \{ 0,1,2,3,4,5\}$ denotes the spectral pixel, $\Gamma \in \{A,B,C,D\}$ denotes the ABCD-channel, $i=0$ stands for the white-light pixel, and $i=1,..,5$ stands for the five spectral pixels. Typical computation rates for fringe tracking are 250~Hz--2~kHz. A number of parameters have to be pre-computed during the calibration and stored in the FSU database, where they can be accessed from the real-time computers. These calibration parameters are:
\begin{enumerate} 
\item \textbf{Dark:} One bias value $G_{i,\Gamma}$ for each pixel used for dark-correction.
\item \textbf{Flat:} One flat-value $F_{i,\Gamma}$ for each pixel used for photometric correction.
\item \textbf{Wavelength:} One effective wavelength value $\lambda_{i,\Gamma}$ for each pixel.
\item \textbf{Phase-shift error coefficients:} 6 values $\alpha_{i}$, $\beta_{i}$, $\gamma_{i}$, and $\delta_{i}$ for each ABCD-channel. 
\item \textbf{Visibility noise:} One value $\upsilon_0$ for the white-light band used to compute the SNR.
\end{enumerate} 
The current algorithms do not take the individual per-pixel wavelengths into account, but the mean wavelength $\lambda_i$ and the corresponding wavenumber $\sigma_i$ in each spectral channel:
\begin{equation}\label{eq:effWl}
	\lambda_i = \frac{1}{4} \sum_\Gamma \lambda_{i,\Gamma} \; , \; \; \sigma_i = \frac{1}{\lambda_i}.
\end{equation} 
The raw pixel intensities are dark-corrected and normalised with the photometric factor to yield the signals $S_{i,\Gamma}$:
\begin{equation}\label{eq:signal}
	S_{i,\Gamma}(t) = \frac{I_{i,\Gamma}(t) - G_{i,\Gamma}}{F_{i,\Gamma} - 2 G_{i,\Gamma}}.
\end{equation} 
Factor 2 in Eq. \ref{eq:signal} originates in the way the database entries are computed, see Sect. \ref{sec:flat}.\\ 
In the ideal case of even-phase separations of $\pi / 2$ between ideal signals $S_{i,\Gamma}^\prime$, the fringe phase $\phi_i$ is computed from the fringe quadratures, $X_i^\prime = S_{i,A}^\prime - S_{i,C}^\prime$ and $Y_i^\prime = S_{i,B}^\prime - S_{i,D}^\prime$, using the classic ABCD formula\footnote{The two argument function \textit{atan2} is used to reach the $2\pi$ non-ambiguous range.} \citep{Shao1977}:
\begin{equation}\label{eq:ABCD}
	 \phi_i = \tan^{-1} \frac{Y_i^\prime}{X_i^\prime} = \tan^{-1} \frac{S_{i,B}^\prime - S_{i,D}^\prime}{S_{i,A}^\prime - S_{i,C}^\prime}.
\end{equation} 
In the real case, where the phase separations of the signals $S_{i,\Gamma}(t)$ are uneven and different from $\pi / 2$, Eq. \ref{eq:ABCD} needs to be generalised by expressing the fringe quadratures as a function of the non-ideal signals $S_{i,\Gamma}(t)$. To achieve this, we define the non-ideal fringe quadratures, $X_i=S_{i,A}-S_{i,C}$ and $Y_i=S_{i,B}-S_{i,D}$, and find that they can be expressed as linear combinations of $X^\prime_i$ and $Y^\prime_i$. Inverting the corresponding $2\times2$~matrix yields the correct fringe quadratures $X^\prime_i$ and $Y^\prime_i$ as function of $X_i$ and $Y_i$, hence of $S_{i,\Gamma}$:
\begin{equation}\label{phasors}
\begin{array}{l}
	X_{i}(t)^\prime = \left[ (S_{i,A}(t)  - S_{i,C}(t))\, \gamma_i - (S_{i,B}(t)  - S_{i,D}(t))\, \alpha_i \right] c_{sc},\\
	Y_{i}(t)^\prime = \left[(S_{i,B}(t)  - S_{i,D}(t))\, \beta_i - (S_{i,A}(t)  - S_{i,C}(t))\, \delta_i \right] c_{sc},
\end{array}
\end{equation}
where the phase shift error coefficients are defined as
\begin{equation}
\begin{array}{l}
	\alpha_i=\sin \psi_{i,C},\; \beta_i=1+\cos \psi_{i,C},\; \gamma_i=\cos \psi_{i,B}+\cos, \psi_{i,D}\; \\
 \mathrm{and} \;\; \delta_i=-\sin \psi_{i,B}-\sin \psi_{i,D} \label{eq:alpha},
	\end{array}
\end{equation} 
and the phase shift errors $\psi_{i,B}$, $\psi_{i,C}$, and $\psi_{i,D}$ are defined as the deviation of the B, C, and D channel phase shifts from their nominal values of $\frac{1}{2} \pi$, $\pi$, and $\frac{3}{2} \pi$, respectively ($\psi_{i,A} = 0$, by definition of channel A as reference). The constant 
\begin{equation}
c_{sc} =  \left(\beta \gamma - \alpha \delta \right)^{-1}
\end{equation} 
is required for correct visibility estimation.\\
Now we can apply Eq. \ref{eq:ABCD} and compute the FSU phase $\phi$ in radians from the white-light signals with a non-ambiguous range of one wavelength $\lambda_0$:
\begin{equation}
	 \phi(t) =  \phi_0(t) =  \tan^{-1} \frac{Y_{0}^\prime(t)}{X_0^\prime(t)}.
\end{equation} 
The optical path difference $\Omega$ measured with the FSU in units of length is then given by
\begin{equation}\label{eq:opd}
	 \Omega(t) =  \frac{\phi_0(t)}{2 \pi}  \lambda_0.
\end{equation}   
To estimate the group delay, we calculate the discrete Fourier transformation (DFT) ${\mathcal{F}}$ of the fringe quadratures 
\begin{equation}
	{\mathcal{F}}(x,t) =  \sum_{i=1}^{5} \, \left(X_{i}^\prime(t) + j Y_{i}^\prime(t)\right) \, \exp{{-j 2 \pi \sigma_i x}}, 
\end{equation} 
where $j=\sqrt{-1}$ and $x$ spans an evenly spaced range of $\pm12 \, \mu$m with $60$ points, i.e. $x = x_1,..,x_N$ with $N=60$, $x_1=-12 \, \mu$m and $x_{60}=12 \, \mu$m. A first estimate of the group delay $\mathrm{GD}$ is the displacement $x_M$ for which the DFT modulus is maximum \citep{Colavita1999a}:
\begin{equation}
	\mathrm{GD}(t) = x_M \hspace{0.5 cm} \mathrm{such\; that} \hspace{0.5 cm}  | {\mathcal{F}}(x_M,t) | \geq  |{\mathcal{F}}(x,t)| \; \; \forall x. 
\end{equation} 
To refine this estimate, we compute the final group delay value from the maximum of the parabolic fit to $| {\mathcal{F}}(x_M,t) |$ and two neighbouring values $| {\mathcal{F}}(x_{M\pm1},t) |$. Consequently, the FSU delivers group delay estimates in units of length over a range of $\pm 12\, \mu$m across the central fringe (Fig. \ref{fig:PDGD}). The visibility amplitude V in white light writes as
\begin{equation}\label{eq:vis}
	\mathrm{V}(t) = \sqrt{\, X_{0}^{\prime 2}(t)  + Y_{0}^{\prime2}(t)}.
\end{equation} 
Finally, the SNR is computed as the ratio of visibility and the visibility noise $\upsilon_{0}$, derived by applying Eq. \ref{eq:vis} to calibration data in absence of fringes (Eq. \ref{eq:vnoise}, see Sect. \ref{sec:fringe}): 
\begin{equation}\label{eq:snr}
	\mathrm{SNR}_{0}(t) = \frac{\mathrm{V}(t)}{\upsilon_{0}}.
	\end{equation}
In summary, the three FSU real-time estimates used by the OPD controller for fringe tracking are the phase in radian, the group delay in meters, and the unitless SNR. The state machine of the OPD controller relies on the $\mathrm{SNR}_0(t)$ in the white-light pixel. 
\section{FSU operation}
\subsection{Laboratory calibration}\label{seq:labcal}
The FSU calibration procedure was developed in the testbed \citep{Sahlmann2008b}. It relies on the PRIMA laser metrology and makes use of the RMN recording facility \citep{Abuter2008}. FSUA and FSUB need to be calibrated independently and the procedure is identical in either case. It is possible to perform FSUA and FSUB calibration simultaneously. The initial step is to inject two beams from the VLTI calibration source (Sect. \ref{sec:calibmet}), powered by a black body cavity at $700\, ^\circ$C, into the fringe sensor. The beam propagation distance from the calibration source to the FSU entrance is $\sim$9~m, and the metrology laser beams travel it in doublepass.\\
Before calibration, the beam injection is optimised by circular modulation of M4 and feeding back the alignment error derived from synchronous de-modulation of the sum of the four white-light intensities $\sum_{\Gamma} I_{0,\Gamma}(t)$. This procedure is adapted from FINITO \citep{Bonnet2006}.
\begin{figure}
\includegraphics[width= \linewidth]{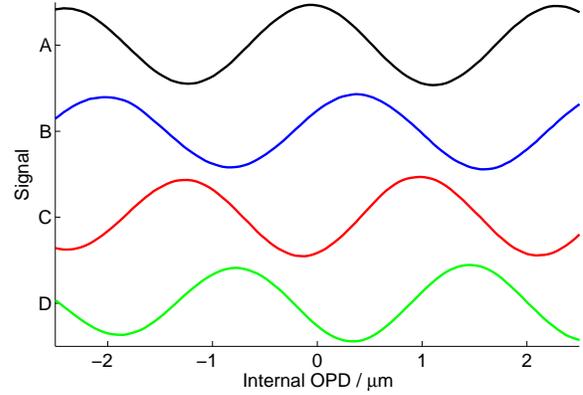} 
\caption{White light fringes close to zero OPD from a calibration scan. The signals are vertically offset for clarity.}
\label{fig:fringes}
\end{figure}
\subsubsection{Dark calibration}\label{sec:dark}
The bias values $G_{i,\Gamma}$ for the dark-correction are computed from a 10~s record of the detector signals $G_{i,\Gamma}(t)$ in the 24 pixels, where both incoming beams are off-pointed by the FSU alignment system. The dark-calibration template computes the mean values $G_{i,\Gamma} = \langle G_{i,\Gamma}(t) \rangle $, where the brackets denote time averaging and updates the database entries accordingly.
\subsubsection{Flat calibration}\label{sec:flat}
The photometric flat-signals $F_{i,\Gamma}$ are obtained in two steps. First the beam B2 (Fig. \ref{fig:fsu_principle}) is off-pointed and a 10~s record of the 24 pixels is taken, yielding the flat signals $F1_{i,\Gamma}(t)$ of the beam B1. The analogous procedure provides the flat signals $F2_{i,\Gamma}(t)$ of beam B2. The flat calibration template updates the FSU database with the flat values $F_{i,\Gamma} = \langle F1_{i,\Gamma}(t) \rangle  + \langle F2_{i,\Gamma}(t) \rangle$. The dark-correction of these values is ensured with Eq. \ref{eq:signal}. 
\subsubsection{Fringe calibration}\label{sec:fringe}  
\begin{figure}
\includegraphics[width= \linewidth]{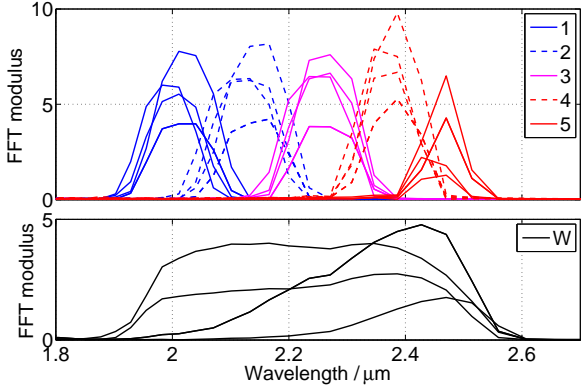}
\caption{Fourier Transform modulus of all 24 pixels from a calibration run, corrected for the black body spectrum. The wavelength variation in the spectral pixels (1-5, {\it top}) and the broad-band white-light pixel (W, {\it bottom}) are visible.}
\label{fig:fft}
\end{figure}

The computation of phase, group delay, and SNR requires the precise values of the effective wavelength and phase shift error of each pixel and the visibility noise in the white-light band. These parameters are deduced from the fringe calibration. We use Fourier Transform Spectroscopy to derive the effective wavelengths. Several OPD scans over the white-light fringe packet are performed with a linear motor of the alignment system, while the FSU pixel intensities and the internal OPD, measured with the laser metrology system, are recorded simultaneously. For each of the 24 pixels, we combine the consecutive scans based on the metrology reference. The fringes are smoothed and interpolated for even spacing (Fig. \ref{fig:fringes}) before computing the effective wavelengths $\lambda_{i,\Gamma}$ from the barycentre of the Fourier transform modulus (Fig. \ref{fig:fft}). The default calibration parameters are four fringe scans and an effective scanning range of 140~$\mu$m. This leads approximately to 50 samples per fringe and a wavelength resolution of 32~nm. The effective wavelengths of ABCD pixels typically differ by some percent. The relative phase shifts of the ABCD-channels are derived by cross-correlating their fringe packets. Offsets of the crosscorrelation-functions with respect to the autocorrelation of channel A are converted into phase, based on the effective wavelengths. The resolution here is 1~mrad. Eventually, the phase shift error coefficients $\alpha_{i}$, $\beta_{i}$, $\gamma_{i}$, and $\delta_{i}$ are derived from Eq. \ref{eq:alpha}. \\
We measure large deviations of the phase shifts from the nominal values of $\frac{1}{2} \pi$, $\pi$ and $\frac{3}{2} \pi$, which can reach $0.25 \, \pi$ in extreme cases, and a slight chromatic dependence. The reasons for these large deviations are imperfections in the FSU optical components, especially in the achromatic retarder and the beam combiner, and in relay optics of the VLTI laboratory. By design, the phase difference between p- and s-polarisation generated by the phase shifter is $(\frac{1}{2} \pm 0.002) \, \pi$ across \textit{K}-band. The beam combiner semi-reflective coatings introduce phase differences below $\pm \, 0.05 \,\pi$. Polarising beam-splitter leakage associated with beam combiner differential phase retardance accounts for $0.11 \, \pi$, whereas the observed values can be more than twice this number. Therefore the observed deviations can only be explained partially by the known warm optics characteristics. This may stem from the unknown contribution to differential phase retardance of the relay mirrors between the calibration source and the FSU.\\
The calibration procedure corrects for the deviations and makes it possible to recover phase and group delay. However, non-ideal phase shifts increase the phase noise and deteriorate group delay and SNR estimates (Sect. \ref{sec:RTest}).\\
Once the wavelengths and phase shift errors are known, the visibility noise $\upsilon_0$ can be computed from the normalised flat exposure $\hat{S}_{0,\Gamma}(t)$:
\begin{equation}
	\hat{S}_{0,\Gamma}(t) = \frac{(F1_{0,\Gamma}(t) - G_{0,\Gamma}) + (F2_{0,\Gamma}(t) - G_{0,\Gamma})}{F_{0,\Gamma} - 2 G_{i,\Gamma}},
\end{equation} 
\begin{eqnarray}\label{eq:vnoise}
	\upsilon_{0}^2 & = \langle [(\hat S_{0,A}(t)  - \hat S_{0,C}(t)) \, \gamma_0 - (\hat S_{0,B}(t)  - \hat S_{0,D}(t)) \, \alpha_0]^2  \nonumber \\
 & + [(\hat S_{0,B}(t)  - \hat S_{0,D}(t)) \, \beta_0 - (\hat S_{0,A}(t)  - \hat S_{0,C}(t)) \, \delta_0]^2\rangle \, c_{sc}^2. % \label{eq:vnoise}
\end{eqnarray} 
\subsection{Real-time estimates}\label{sec:RTest}
\begin{figure}
\includegraphics[width= \linewidth]{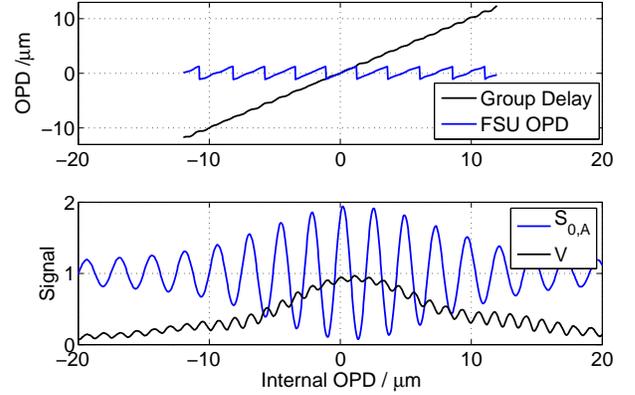} 
\caption{{\it Top}: FSU OPD (from Eq. \ref{eq:opd}, blue) and group delay estimates (black) obtained on a calibration scan. {\it Bottom}:  Visibility V (black) and one FSU signal (blue) as function of internal OPD. Group delay is delivered over a range of $\pm 12 \mu$m and the FSU OPD is wrapping with a $\lambda_0$-period. The coherence length of the bandpass is roughly 30~$\mu$m.}
\label{fig:PDGD}
\end{figure}

The linearity of phase and group delay estimates is essential for guaranteeing a reasonable control signal for the fringe tracking loop. From the OPD scans during calibration, the non-linearity is estimated as the deviation of the delay measurement slope from unity, where the laser metrology measurement serves as reference (Fig. \ref{fig:PDGD}). The phase non-linearity is within specification below 10~\% over the central fringe, while the group delay non-linearity exceeds the specified 20~\% over the central $\pm$ 6 $\mu$m by a factor of 3, which comes from the real-time algorithm not taking the individual wavelengths of the ABCD-channels into account, but instead their mean value (Eq. \ref{eq:effWl}). The resulting wavelength mismatch causes a periodic error with double frequency, an effect also visible in the visibility amplitude plotted in Fig. \ref{fig:PDGD}. However, the gain margin of the fringe-tracking control loop amounts to approximately 10~dB and is large enough to cope with the phase non-linearity.\\
In the testbed, the 3dB-bandwidth of the phase estimate was measured to $(910 \pm 40)$~Hz with an integration time of 0.5~ms. In addition, we measured the phase noise performance, which sets the FSU sensitivity, hence the fringe-tracking limiting magnitude. For detector integration times (DIT) of $0.5$ ms, 1~ms, and 2~ms with readout-noise of $(27 \pm 1)$, $(19 \pm 1)$, and $(18 \pm 1)$ electrons RMS per pixel, respectively, we found the phase RMS noise to be 1.1 to 1.6 times greater than the theoretical value \citep{Sahlmann2008a}, resulting in a theoretical sensitivity loss of $\sim$0.5 magnitudes.   
\subsection{Night calibration}\label{sec:ncal}
Every time the FSU acquires a new stellar object, the night calibration is executed. It consists of measuring the sky background and the object's photometry, following the same procedure as described for the laboratory dark (Sect. \ref{sec:dark}) and flat (Sect. \ref{sec:flat}) calibration, respectively, and re-computing the visibility noise (Eq. \ref{eq:vnoise}). As a result the FSU database is updated with the new background, photometric, and visibility noise values, which are henceforth considered for the real-time computations.\\
At a later stage, when both FSUA and FSUB are available with the star separator modules, it is possible to perform the fringe calibration on the observed object, to account for the spectral transmission of the VLTI beamtrain and the object's spectrum. The procedure becomes more elaborate because of the telescopes, delay lines, and other VLTI-subsystems, but in principle one fringe sensor will be used to stabilise the fringes, while the other fringe sensor performs the fringe calibration as described in Sect. \ref{sec:fringe}.
\begin{figure}
	\includegraphics[width= \linewidth]{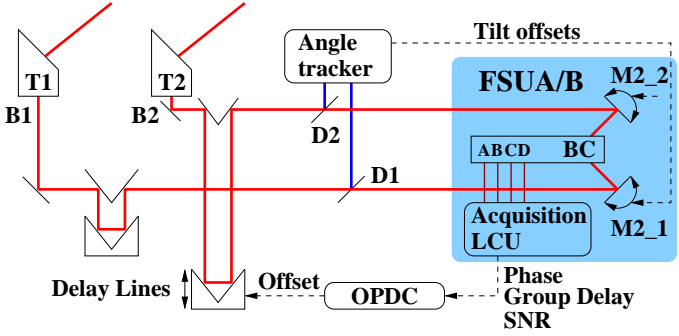}
	\caption{Simplified VLTI control system for single-feed fringe tracking with FSUA or FSUB. Telescopes T1 and T2 capture the starlight, and beams B1 and B2 are routed through the main delay lines. Dichroic mirrors D1 and D2 feed the \textit{H}-band light to the angle tracking camera sending open-loop tilt offsets to the piezo-controlled mirrors M2\_1 and M2\_2, as part of the FSU. \textit{K}-band beams are combined (BC) and the ABCD signals are used to compute phase, group delay, and SNR. These estimates are used by the OPD controller (OPDC) to offset one delay line for closed loop fringe tracking.}
	\label{fig:vlti}
\end{figure}
\section{FSU fringe tracking} \label{sec:FSUftk}
The presented results are obtained in single-feed, on-axis fringe tracking. For our test purposes, the role of the first VLTI fringe tracking sensor FINITO was taken over by either FSUA or FSUB, but limited to two telescope observations, and the control system as described by \cite{Haguenauer2008} and \cite{Bouquin2008} remained nearly identical. In this section we briefly describe the VLTI control system and highlight the differences when observing with the FSU instead of FINITO.\\
Figure \ref{fig:vlti} gives a simplified overview of the interferometer control system, when used in conjunction with either FSUA or FSUB. The target is observed with two ATs or two UTs, providing field stabilisation with a tip-tilt system \citep{Koehler:2006rr} or adaptive optics \citep{Arsenault:2004cr}, respectively. Mirrors guide the telescope beams through optical delay lines into the beam combination laboratory, where the \textit{H}-band beams are fed into the angle-tracking camera \citep{Gitton:2004wd} and the \textit{K}-band beams enter the FSU. When the \textit{H}-band fringe sensor FINITO is used, the \textit{K}-band is used for angle tracking. The angle-tracking camera computes tip-tilt error signals with respect to the alignment reference position at the read-out rate of typically 100-300 Hz. Slow angle tracking (bandwidth $\sim$1~Hz, limited by communication delays) is performed in closed loop using actuators at the telescopes. If the target is bright enough for fast angle tracking, the high-frequency error signals are forwarded in open loop to the FSU M2 tip-tilt stages (described for FINITO by \citealt{Bonnet2006}). The feed-forward is disabled as soon as the camera has to run at read-out rates below 100~Hz, which from our experience is required for target magnitudes fainter than $m_H=6$, to avoid the amplifying regime of the corresponding rejection function.\\ 
The FSU real-time estimates of phase, group delay, and SNR are picked up by the OPD controller, which sends delay offsets to one delay line, whereas the other delay line remains fixed. The delay offset includes the pre-computed trajectory based on the interferometer configuration and the target coordinates and the fringe tracking offset computed in real time from the fringe sensor estimates.\\
The three-state OPD controller used for our observations samples at 2~kHz and is described by \cite{Bouquin2008}. Three user-defined SNR thresholds (\textit{det}, \textit{close}, and \textit{open}) define transitions between the three states 
\begin{description}
\item[SEARCH] A triangular search trajectory is performed around the predicted fringe position. Fringe tracking loop is open.
\item[IDLE] The delay line follows the predicted fringe position. Fringe tracking loop is frozen. 
\item[TRACK] Fringe tracking offsets are sent to the delay line, based on the fringe sensor estimates. Fringe tracking loop is closed.
\end{description}
as explained in Table \ref{tab:opdc}. The fringe-tracking control algorithm used by the OPD controller with FSU estimates combines both slow group delay and fast phase tracking. The phase estimate is unwrapped if the difference between two consecutive samples is more than $\pi$. Instead of controlling to zero-phase, the phase controller tracks a time varying target designed to maintain the group delay signal at zero. This variable target depends on the integral of the group delay. The time constant of the group delay integral controller is a few seconds and large compared to the $\sim$0.1~s phase controller time constant.  In consequence, the control algorithm is tracking the delay of maximum coherence within the central fringe. The gains of this controller were defined during the testbed phase by fringe tracking on a model atmosphere \citep{Sahlmann2008a} and are found to match for on-sky observations.
\begin{table} 
\caption{OPD controller state transition table (adapted from \citealt{Bouquin2008})}
\label{tab:opdc} 
\centering  
\begin{tabular}{r r | c c c } 
\hline\hline % 
			& next 			& SEARCH 						& IDLE 						& TRACK  \\
current 	&   				&  									&  								&   \\ 
\hline 
			&   				&  									&  								&   \\ 
SEARCH 	& 				& $\mathrm{SNR} < {det}$ 		& - 								& $\mathrm{SNR} > det$  \\ 
			&   				&  									&  								&   \\ 
%IDLE  		& 				& $\mathrm{SNR} < open$ 		&  	$\mathrm{SNR} < det$	& $\mathrm{SNR} > det$  \\ 
%			&				& for 20 ms						&								& \\
%			&   				&  									&  								&   \\ 
%TRACK 	& 				& $\mathrm{SNR} < open$ 		& $\mathrm{SNR} < det$ 	& $\mathrm{SNR} > det$ \\ 
%			& 				&  									& and $\mathrm{SNR} > open$& \\ 
IDLE  		& 				& $\mathrm{SNR} < close$ 		&  	$\mathrm{SNR} < close$	& $\mathrm{SNR} > close$  \\ 
			&				& for 20 ms						&								& \\
			&   				&  									&  								&   \\ 
TRACK 	& 				& -  		& $\mathrm{SNR} < open$ 	& $\mathrm{SNR} > open$ \\ 
			& 				&  									& & \\ 
\hline  
\end{tabular} 
\end{table}
\subsection{Installation, alignment, and first fringes}\label{sec:firstfringes}
\begin{figure}
\includegraphics[width= \linewidth]{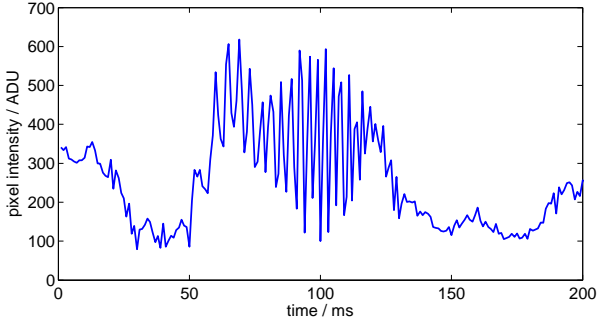}
\caption{First FSUA fringes obtained on 3 Sep. 2008 while observing \object{HD19349} ($m_K=0.4$) and scanning the OPD with one linear motor of the alignment system. Raw intensity differences $I_{0,A}(t) - I_{0,C}(t)$ of the white-light pixels in quadrants A and C are shown. The integration time is 1~ms.}
\label{fig:FSUAFringes}
\end{figure}

As part of the PRIMA facility, the FSU was installed at the VLTI during a seven-week period starting in July 2008. First stand-alone tests were possible after placing the opto-mechanical systems and installing the computer infrastructure. One important milestone was the first laboratory calibration, because it involves several VLTI subsystems and requires the calibration source, the laser metrology, and the reflective memory network to be functional. Subsequently, the FSU was aligned within the VLTI laboratory reference frame and interface tests to the angle tracking system were successfully carried out. Finally, communication with the main delay lines for fringe tracking was established via the OPD controller, which made FSUA ready to go on sky. FSUB became operational in January 2009, after correction of an initial cold camera defect. Another milestone was achieved when FSUA recorded first fringes with two ATs (Fig.~\ref{fig:FSUAFringes}).\\
Initial FSUA tests were performed with a cold camera configuration where half of the detected light falls in the white-light pixel and the other half is distributed on the five spectral pixels (see Sect.~\ref{sec:lrs}), the flux ratio for the white-light pixel $T_W$ and for the spectral pixels $T_S$ each equal 50~\%. We denote this configuration as FSUA[0].\\
In January 2009 both FSUA and FSUB cold cameras had to be modified such that all light falls on the spectral pixels. This became necessary because of difficulties during the alignment of the prism assembly, causing substantial efficiency loss (cf. Table~\ref{table:ccc}), and because of degrading coatings on the $\mathrm{BaF}_2$-prisms, making these prisms unusable. The white light pixel intensity $I_{0,\Gamma}(t)$ is then replaced in software by the sum of the spectral pixel intensities to provide a synthetic white light intensity 
\begin{equation}
\hat{I}_{0,\Gamma}(t) = \sum_{i=1}^{5} I_{i,\Gamma}(t).
\end{equation}
These configurations are referred to as FSUA[1] and FSUB[1] ($T_W = 0$~\% and $T_S=100$~\%). Table \ref{table:ccc} summarises the configurations. The light loss is computed from the amount of light that falls on adjacent pixels due to imperfect optics and is not detected. $T_W$ and $T_S$ are given with respect to the detected light. Uncertainties of these numbers are smaller than 1~\%.
\begin{table} 
\caption{FSU cold camera configurations} 
\label{table:ccc} 
\centering  
\begin{tabular}{c r r r } 
\hline\hline % 
Configuration & $T_W$ & $T_S$ & Loss  \\
   & (\%) &  (\%) &  (\%)  \\ 
\hline 
FSUA[0] & 49 & 51 & 27  \\ 
FSUA[1] & 0 & 100 & 13  \\ 
FSUB[1] & 0 & 100 & 5  \\ 
\hline  
\end{tabular} 
\end{table} 
\subsection{Data recording and exploration}
Data recording during FSU operation, both on sky and for the laboratory calibration, relies exclusively on the RMN recording facility \citep{Abuter2008}. It allows us to record virtually all data available on the RMN in real time at kHz-rates and has previously been implemented at the VLTI \citep{Bouquin2009b}. All real-time computers involved in the fringe-tracking control system communicate via the RMN and all significant data can be captured with this facility. Based on timestamps delivered with each data product and recorded in the produced file of one exposure, it is possible to correlate each sample of real-time data from different computers during the data analysis. For instance, if FSUA is sampling at 1~kHz and the OPD controller runs at 2~kHz, it is possible to identify the controller state for every FSUA estimate. As a result, it is straightforward to clean FSUA data from periods where the fringe tracking loop is not closed. This results in an improvement of the FSUA data quality. The same principle applies to the angle-tracking system or later on within PRIMA for the differential OPD controller.\\
PRIMA, once fully operational, will equally rely on the same data recording facility and be able to profit from these data-quality improvement possibilities.  
\section{Results}
\subsection{Fringe tracking with ATs}
\begin{figure*}
\begin{center} 
\includegraphics[width= \linewidth]{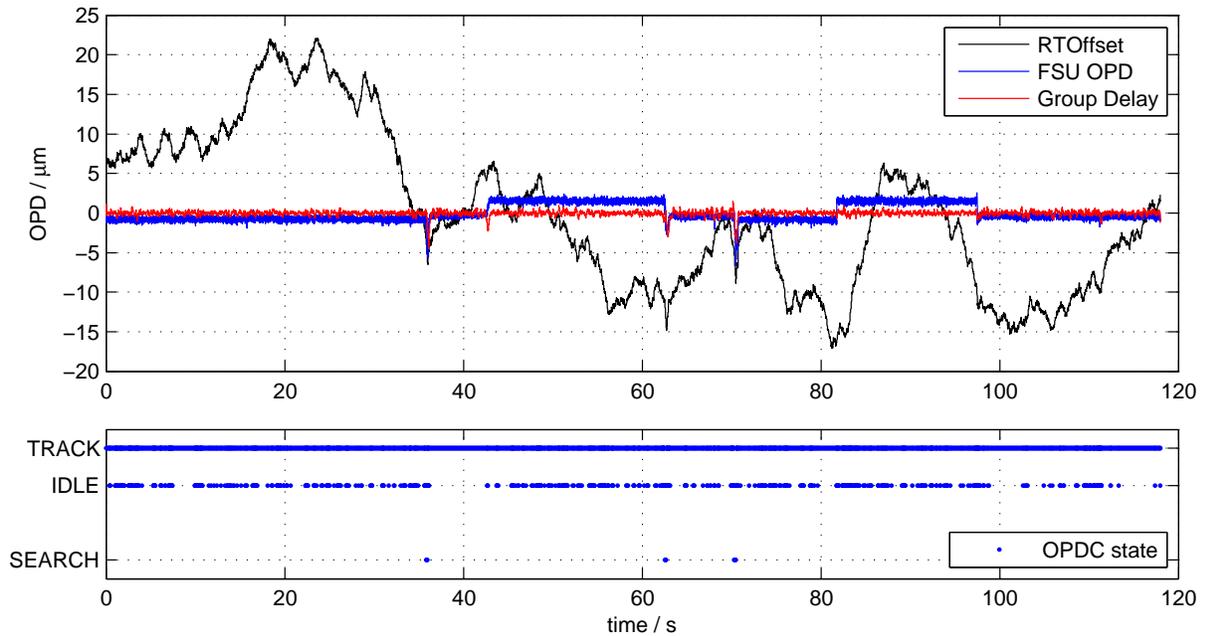}
\end{center}
\caption{FSUA fringe-tracking sequence observing HD4803 ($m_K=7.6$) on 25 Nov. 2008. {\it Top}: FSU OPD (blue), low-pass filtered group delay (red), and the real-time offset (RTOffset) sent to the tracking delay line (black), which is a filtered representation of the atmospheric piston, in $\mu$m. FSU OPD is obtained by unwrapping the phase and applying Eq. \ref{eq:opd}. Group delay is controlled to zero and different FSU OPD levels are visible because the occurring fringe jumps cannot be corrected for by the unwrapping algorithm. {\it Bottom}: OPD controller state. This is achieved at a sampling rate of 500~Hz, a baseline length of 64~m with two ATs, and in good atmospheric conditions with $0.7\, \arcsec$ seeing and 5.0~ms coherence time. The lock ratio is 99~\% and the residual OPD is $\sim$190~nm RMS.}
\label{fig:FTK}
\end{figure*}

Two nights after having recorded first fringes, the VLTI fringe tracking loop was closed with FSUA as sensor. In the subsequent nights, the FSU acquired and tracked fringes down to a magnitude of $m_K=5.7$\footnote{Stellar magnitudes are taken from Centre de Donn{\'e}es astronomiques de Strasbourg (CDS).}. Several baselines with lengths of 32~m, 72~m, and 96~m were used, while making pairwise use of all four available ATs. To avoid nebulosity, we define fringe tracking as achieving a fringe lock ratio above 70 \% over one minute, whereas there is no consideration of the tracking residuals.\\
During the first PRIMA commissioning runs in October and November 2008, fringe tracking was routinely performed with FSUA with baselines in the range of $32\--96$~m and two ATs, respectively. For the February 2009 commissioning the cold cameras of FSUA and FSUB were modified as described in Sect. \ref{sec:firstfringes} and FSUB performed first fringe tracking.\\
Tracking robustness was demonstrated for more than 100 mostly unresolved stars at zenith angles of less than 40~$\deg$ and for various atmospheric conditions and target magnitudes. Within 30 commissioning nights, it was possible to sample the magnitude space from $m_K\sim0\--10$ in atmospheric conditions ranging from poor to excellent (seeing $\sim0.4\--2.2 \, \arcsec$ , coherence time $\tau_0\sim8\--0.8$~ms). Atmospheric conditions were retrieved from the Paranal observatory seeing monitor operating at visible wavelength. Depending on these parameters, FSUA or FSUB was capable of locking fringes at sampling rates between 100~Hz and  2~kHz. By the end of the third commissioning run, FSUA at 250~Hz had achieved fringe tracking with two 1.8~m ATs on an $m_K=9.0$ star, and fringes were recorded but not tracked with FSUB at 100~Hz on a star of $m_K=10.0$ (Table \ref{table:1}). Figure \ref{fig:FTK} illustrates a fringe-tracking sequence. 
\begin{table*} 
\caption{Examples of FSU fringe tracking with ATs during commissioning} 
\label{table:1} 
\centering 
\begin{tabular}{r c l r c c c c c} 
\hline\hline 
Date & Configuration & Target & $m_K$ & Baseline & Seeing & $\tau_0$ & DIT  & Lock ratio \\ 
 &  &  & &  (m) &  ($\arcsec$) & (ms)  & (ms) &  (\%) \\  
\hline  
8 Sep. 2008 & FSUA[0] & \object{HD206647} & 5.7 & 96 & 0.7 & 3.2 & 2 & $>$ 80 \\  
21 Oct. 2008 & FSUA[0] & \object{HD10067} & 7.6 & 32 & 0.5 & 3.5 & 2 & $>$ 70 \\
25 Nov. 2008 & FSUA[0] & \object{HD4803} & 7.6 & 64 & 0.7 & 5.0 & 2 & $>$ 95 \\
28 Nov. 2008 & FSUA[0] & \object{HD18558} & 8.3 & 96 & 0.7 & 3.0 & 4 & $>$ 70 \\
28 Nov. 2008 & FSUA[0] & \object{HD17967} & 9.0 & 96 & 0.7 & 3.0 & 4 & $>$ 70 \\
28 Nov. 2008 & FSUA[0] & \object{HD23747} & 9.5 & 96 & 1.0 & 2.3 & 4 & $\sim$ 2 \\
4 Feb. 2009 & FSUB[1] & \object{HD100091} & 8.6 & 48 & 0.8 & 4.6 & 4 & $>$ 80 \\
5 Feb. 2009 & FSUA[1] & \object{HD100091} & 8.6 & 48 & 0.6 & 5.7 & 10 & $>$ 70 \\
5 Feb. 2009 & FSUB[1] & \object{[SMO84] 131534.25-330000.8} & 10.0 & 48 & 0.6 & 7.0 & 10 & $\sim$ 10 \\
\hline  
\end{tabular} 
\end{table*} 
\subsection{Data and estimators}
During the 30 commissioning nights, more than 1500 files in standard ESO \textit{fits}-format were recorded with a data volume of 50 Gigabyte. Roughly one third corresponded to fringe tracking observations, while the remaining files were distributed between laboratory and night calibrations. The particularity of FSU / PRIMA raw files is that they contain a data stream with samples at kHz rates. This increases the file size but offers the opportunity to reduce the complete dataset and gives access to high-frequency measurements. We wrote \texttt{Matlab} code to reduce this data, extract the significant parameters, and collect them in a database, allowing us to access the complete sample by defining selection criteria. Basic parameters of the observed targets are extracted from the Simbad astronomical database.\\
Parameters with high dynamic variability, e.g. tracking residuals and lock ratios, are typically computed in a running window of fixed length over the data file. Characterising values of the obtained distribution (e.g. median, mean, RMS) are the results stored in the database. A file selection based on quality estimators was performed before running the reduction, and the number of files is indicated for the results.
\begin{figure}
\begin{center} 
\includegraphics[width= \linewidth]{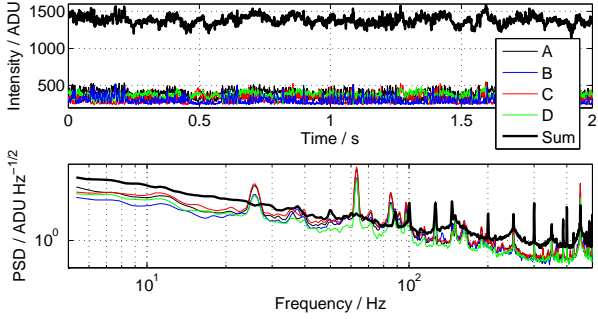} \end{center} 
\caption{White-light intensities during fringe tracking in normal conditions ($\tau_0 = 3.4$~ms, observation run identical to Fig. \ref{fig:opdpsd}). \textit{Top}: Time series of the four quadrants and their sum. The RMS of the sum is 80~ADU. \textit{Bottom}: Corresponding spectra showing strong features at 26 and 63 Hz.}
\label{fig:fluxpsd}
\end{figure} 
\subsection{Injection}\label{sec:inj}
\begin{figure}
\begin{center} 
\includegraphics[width= \linewidth]{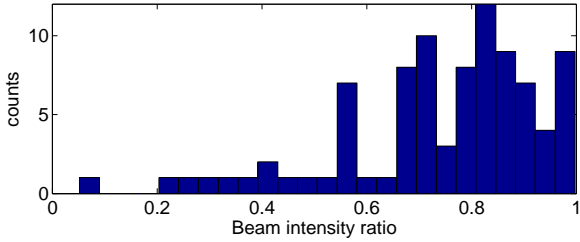} \end{center} 
\caption{Beam intensity ratio histogram obtained from the night calibrations (90 files). The beam with higher intensity is taken as reference, and a value of 1 indicates balanced beams.}
\label{fig:beamratio}
\end{figure}

In addition to the effects common to every instrument employing single-mode fibres (flux variations due to uncorrected tip-tilt jitter), the spatial modulation scheme of the FSU imposes more complications. Because the four ABCD beams are generated before injection into the fibres, any differential static or dynamic aberration will cause differential injection, hence falsify the delay estimates. Therefore, good fibre co-alignment is crucial (cf. \ref{sec:bc}).\\
Because the system, including coupling doublets, is not aberration-free, residual tip-tilt jitter creates artificial phase estimates in the FSU, which cannot be distinguished from real phase due to piston. This creates secondary effects during fringe tracking: when the FSU detects artificial phase variations due to differential injection, the OPD controller will inject these perturbations into the optical path via the delay line actuator, which in turn is seen by the FSU. Hence, it is not easy to distinguish between tip-tilt and piston effects with the FSU during fringe tracking.\\
Figure \ref{fig:fluxpsd} shows a typical injection sequence and illustrates the problem, together with Fig. \ref{fig:opdpsd}. Wavefront perturbations in the telescope beams cause fast injection variability and make the intensity sum vary by 5~\% RMS. Spectra of the ABCD intensities reveal strong perturbations at 26 and 63~Hz, which are also present in the piston spectrum, whereas their sum is flat at these frequencies, therefore these perturbations have to originate in piston vibrations and not from residual tip-tilt. On the other hand, the features around 100 and 126 Hz appear in both individual and summed intensities and are thus at least partially caused by residual tip-tilt. Typical average coupling losses due to residual tip-tilt are in the range of 30 - 40~\%. In poor conditions, the equivalent figures look different and flux dropouts lasting several ms inhibit efficient fringe tracking, which is reflected in Fig. \ref{fig:lr}.\\
In addition to the dynamic intensity imbalance of the input beams, caused by uncorrelated tip-tilt jitter, there is also the quasi-static intensity imbalance that remains approximately constant during one observation and is caused by differences in transmission and alignment of the beamtrains. Its magnitude can be estimated from the photometric night calibration executed before the observation. Figure \ref{fig:beamratio} shows the beam intensity ratio for a set of observations. In $\sim$80~\% of the cases, the intensity ratio is greater than 0.65, and the resulting reduction in visibility amplitude is below 3~\%.
\subsection{Transmission}
\begin{figure}
\begin{center} 
\includegraphics[width= \linewidth]{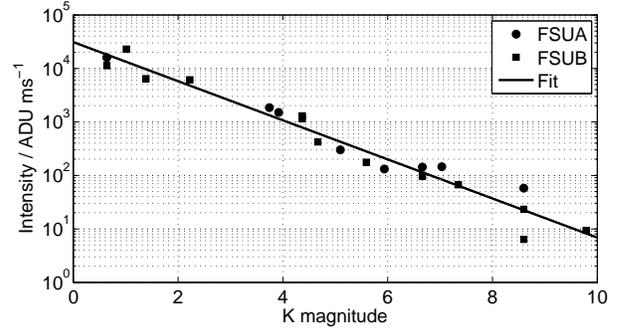} \end{center}   
\caption{FSU total intensity as function of magnitude (22 files). The detected intensity is the calculated from the sum of all pixel intensities.}
\label{fig:transm}
\end{figure}

From the FSU photometric measurements collected during fringe tracking, we can estimate the effective transmission of the instrument. Figure \ref{fig:transm} shows the total intensity detected in FSUA or FSUB after the cryostat intervention. From the exponential fit and the expected incident flux, we estimate the effective transmission to $11\pm3$~\% in \textit{K}-band. This number includes the detector quantum efficiency, both static and dynamic coupling losses, and the cold camera losses (Table \ref{table:ccc}). The theoretical incident intensity is computed assuming a VLTI \textit{K}-band transmission of $35$~\% \citep{gitton2009} and neglecting atmospheric extinction. For the selected sample, the average dynamic coupling loss is $37\pm5$~\%, which is a coarse estimation based on the ratio of maximum and mean injected flux. With this number we can obtain the FSU effective optical transmission of $18\pm5$~\%, including the detector quantum efficiency.
\subsection{Lock ratio}
\begin{figure}
\begin{center} 
\includegraphics[width= \linewidth]{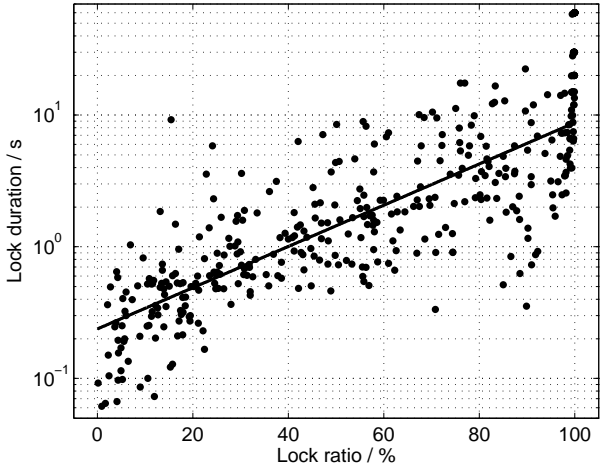} \end{center}  
\caption{Lock duration over one minute as function of lock ratio. The exponential fit is shown (437 files).}
\label{fig:ld}
\end{figure}

\begin{figure*}
\begin{center} 
\includegraphics[width= \linewidth, bb= -153   246   796   566]{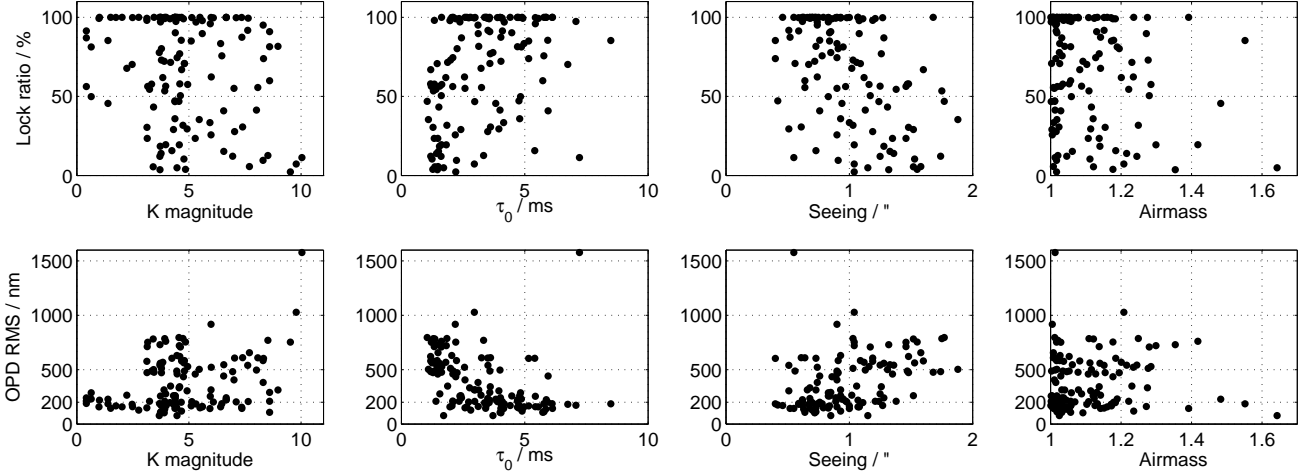} \end{center} 
\caption{Lock ratio (\textit{top row}) and residual OPD (\textit{bottom row}) as function of \textit{K}-band magnitude, coherence time, seeing, and airmass (106 files). Coherence time and seeing are measured in the visible by the seeing monitor.}
\label{fig:lr}
\end{figure*}

\begin{figure}
\begin{center} 
\includegraphics[width= \linewidth]{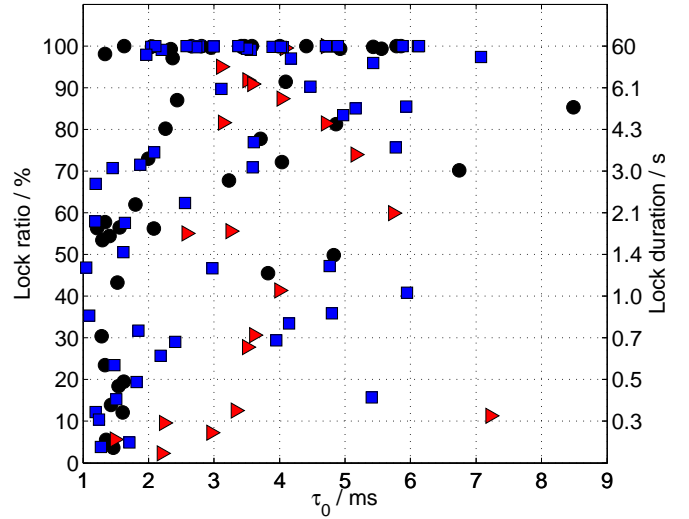} \end{center} 
\caption{Lock ratio as function of visible coherence time (106 files). The righthand ordinate axis shows the average lock duration deduced from the fit of Fig. \ref{fig:ld}. Three target magnitude ranges are indicated by black circles ($\mathrm{m}_K<4$), blue squares ($4<\mathrm{m}_K<7$), and red triangles ($7<\mathrm{m}_K$).}
\label{fig:magtau}
\end{figure}

Two basic parameters that characterise fringe tracking are the lock ratio and lock duration. The lock ratio defines the observation efficiency, and the lock duration sets the maximum time that an instrument, attached to the fringe tracker, can use for coherent integration without perturbations by fringe losses. We define the lock ratio as the fraction of time over one minute that the OPD controller is in state TRACK or IDLE. We include the IDLE state, because in our experience the IDLE state rarely lasts the maximum allowed 20~ms (Sect. \ref{sec:FSUftk}) and fringes are either lost or recovered very quickly. Consequently, we also compute other parameters linked to the TRACK state with this convention, in particular the closed loop phase residuals.\\
Figure \ref{fig:ld} shows the mean lock duration over one minute as function of lock ratio. The data can be approximated coarsely by an exponential function, which shows that an average lock duration longer than 2~s is reached for lock ratios above 60~\%.\\       
Figure \ref{fig:lr} shows the lock ratio as function of the object's \textit{K}-band magnitude and environmental parameters. Ratios above 80~\% are reached for target magnitudes down to $m_K=9.0$. High lock ratios are obtained over a wide range of atmospheric conditions with generally increasing values for improving conditions. Fringe tracking is hardly achieved for seeing above $1.3 \arcsec$ and coherence time below 2~ms.\\
To further illustrate the limits of operation, Fig. \ref{fig:magtau} shows lock ratio and average lock duration as function of coherence time and includes the target magnitude range. For $\tau_0 < 2$~ms, fringe tracking is almost impossible even for bright stars ($m_K < 4$), and faint stars ($m_K > 7$) can usually only be reached for $\tau_0 > 3$~ms. 
\subsection{Residual OPD}
\begin{figure}
\begin{center} 
\includegraphics[width= \linewidth]{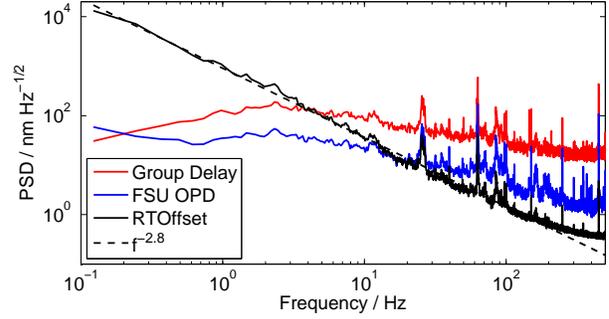} \end{center} 
\caption{Power spectral densities of FSU OPD, group delay, and the delay line RTOffset in normal conditions ($\tau_0 = 3.4$~ms, observation run identical to Fig. \ref{fig:fluxpsd}). A power law with exponent $-2.8$ approximately reproduces the atmospheric signature in the delay line offset.}
\label{fig:opdpsd}
\end{figure}

Another parameter for evaluating fringe-tracking performance is the closed loop residual jitter, which is seen by the fringe tracker or the attached instrument. For PRIMA astrometry, the jitter amplitude (commonly measured in nm RMS over a given time) defines the length of an observation to reach the required accuracy on the differential delay, hence the observing efficiency \citep{Lindegren:1980bu, Shao1992}. For an attached visibility-measuring instrument, the jitter blurs the fringes during the integration and makes the visibility estimation difficult (see e.g. \citealt{Lane:2004rz}, \citealt{Bouquin2008}).\\ 
Both phase and group delay estimates of the FSU can be used to estimate the residual OPD, and we choose the phase for its intrinsically lower noise. In fact, we use the unwrapped phase of the OPD controller in states TRACK and IDLE\footnote{The controller keeps the phase constant when in state IDLE. However, this residual OPD estimator is valid since its dependence on residual group delay is linear.}, which is recorded in the \textit{fits}-files and estimate the residual jitter RMS over 1~s. This means that the closed loop residuals are estimated from the control signal itself and not by an independent measurement. In the case of VLTI, the rejection function is dominated by delays in the communication to the delay line actuator and limits the closed loop bandwidth to $\sim$15~Hz \citep{Lieto2008}.\\
Figure \ref{fig:lr} shows residual OPD as function of stellar magnitude and atmospheric observing conditions. For each night and each target, the file with the highest lock ratio is selected, justified by the OPD controller thresholds being manually adjusted to optimise the fringe tracking (cf. \ref{sec:snr}). Residuals below 200~nm are measured at all magnitudes down to $m_K=8.6$, with a minimum and median value of 80~nm and 280~nm, respectively. A strong correlation between jitter and coherence time is observed and residuals below 300~nm can be expected for $\tau_0 > 4$~ms, whereas they reach 800~nm in poor conditions with $\tau_0 < 2$~ms. Dependence on seeing and airmass is less pronounced, although clearly lower residuals are achieved for seeing $<1.2\arcsec$. The airmass dependence is not fully conclusive due to the low number of points at high airmass, but low residuals are achieved at airmass of $1.7$.\\
Because the FSU detection principle is based on polarisation splitting, the source and beam-train polarisation influence the observations. The expected effect is an increased non-linearity of the FSU response, primarily of the group delay. Since during fringe tracking the group delay is kept close to zero, its non-linearity has a low impact as long as it does not cause the controller track on a non-zero local minimum, which is never observed. However, non-linearity is observed to be problematic in combination with fringe loss events, which is described in Sect. \ref{sec:disp}.\\ 
The actual FSU phase shifts when observing on sky are unknown with the present setup. They can only be measured after dual feed operation is available (see Sect. \ref{sec:ncal}). VLTI does not provide an independent estimate of differential beam-train polarisation, which may also affect the laboratory calibration (cf. \ref{sec:fringe}).\\
The power spectral density (PSD) of residual OPD and group delay, together with the delay line control signal, is shown in Fig. \ref{fig:opdpsd}. The low-frequency atmospheric piston is corrected for by the fringe tracking control loop up to the system's bandwidth of $\sim$15~Hz. The features at 26 and 63~Hz can thus not be cancelled and are identified as pure piston perturbations with the help of the reasoning in Sect. \ref{sec:inj}. They contribute $\sim$30 and $\sim$50 nm RMS to the OPD residual, respectively. They are transient features with unknown origins and they apparently appear randomly in roughly 35~\% of the observations. Most likely they have to be attributed to a vibrating optical element in the VLTI beam-train.\\
The group delay noise floor is one order of magnitude above the OPD noise floor, as expected from its definition as a wavelength derivative of the phase and the applied algorithm (e.g. \citealt{Lawson:2000sf}). At the observatory, an unidentified source of parasitic noise increases the read-out noise to $31\pm4$ electrons RMS per pixel for a detector integration time of 1~ms compared to $19\pm1$ electrons RMS measured in the testbed. However, a considerable fraction of this additional noise is correllated for all quadrants, and the effective noise of a two-pixel difference is $21\pm4$ electrons RMS, which eventually affects the FSU estimates.\\
The atmospheric turbulence shows a slope of approximately $-2.8$, which is the typical value throughout our observations and compatible with the nominal value of $-8/3$ from Kolmogorov theory and the value of $-2.5$ measured at the Palomar testbed interferometer \citep{Lane:2003zl}.\\ 
This fringe tracking data can be used for atmospheric turbulence measurements to derive e.g. the \textit{K}-band coherence time and outer scale size of the VLTI site, as demonstrated by \cite{Linfield:2001rr} for the Palomar testbed interferometer.    
\subsection{Visibility}\label{sec:vis}
\begin{figure}
\begin{center} 
\includegraphics[width= \linewidth]{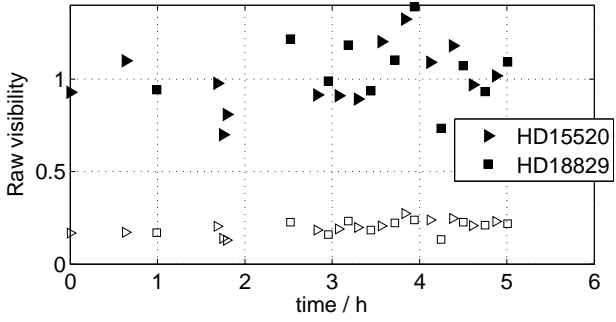} \end{center} 
\caption{FSU raw visibility measurements. Filled symbols indicate measurements in the fringes (controller state TRACK), whereas open symbols are obtained in the absence of fringes (controller state SEARCH). The projected baseline length decreases from 65~m to 50~m during this sequence. Atmospheric conditions are stable with seeing around $1~\arcsec$.}
\label{fig:vis}
\end{figure}

We present FSU raw visibility measurements obtained during one night while repeatedly observing a set of stars with similar coordinates. Each observation consists of acquiring the star, performing the night calibration, and taking 120~s of fringe-tracking data. Thirty observations were carried out within 5~h, which results in one observation every 10~min. This is achieved in single-feed engineering mode and for grouped stars. The raw visibility amplitude is computed from Eq. \ref{eq:vis} for each sample obtained at a 1~kHz rate. Figure \ref{fig:vis} shows alternating observations of two stars. Each point is obtained as the mean value from several thousands of normally distributed visibility estimates, with a resulting standard deviation that is smaller than the symbol size, and is corrected for average beam intensity imbalance obtained from the night calibration. Table \ref{table:vis} shows the results, along with selected target characteristic. Mean visibility errors of each target are estimated from the RMS of the dataset. Both stars are unresolved with visibilities compatible with $1.0$. The large scatter of visibility values on the same target may be attributed to limitations of the system and effects not taken into account by our simple algorithm (e.g. photon noise, cf. \citealt{Colavita1999}) or to changes of the interferometer transfer function, although the findings of \cite{Bouquin2009b} suggest the former possibility.\\
A limitation of the FSU is that dynamic differential injection in the ABCD fibres and the instantaneous photometry for the individual beams cannot be monitored in real-time, and cannot be used to correct the visibility estimates.\\
The visibility bias due to the factors listed above is visible in Fig.~\ref{fig:vis} as non-zero values obtained in the absence of fringes and has the value $0.20\pm0.04$. A preliminary analysis with a similar approach to \citealt{Colavita1999} shows that the bias contribution of read-out and photon noise is $\sim$$0.05$ and not sufficient to explain the observed value. The contributions of differential injection and calibration errors remain to be quantified.\\
As demonstrated by \cite{Bouquin2009b}, VLTI fringe tracking data can be used for high-precision stellar diameter measurements. The results obtained here indicate that this is not possible with the FSU and the current calibration and data reduction strategy. This has no immediate effect on the scientific application, because precise visibility measurement capability is not part of the FSU specifications. 
\begin{table} 
\caption{FSUA raw visibilities}
\label{table:vis}  
\centering  
\begin{tabular}{c  c c }  
\hline\hline 
Target & $m_K$  & V   \\
 %		&   		& 	  \\ 
\hline  
\object{HD15520} & 4.5 	& $1.00 \pm  0.17$ 	\\
\object{HD18829} & 4.2  	& $1.04 \pm 0.17$ 	\\
\hline  
\end{tabular} 
\end{table} 
\subsection{SNR}\label{sec:snr}
\begin{figure}
\begin{center} 
\includegraphics[width= \linewidth]{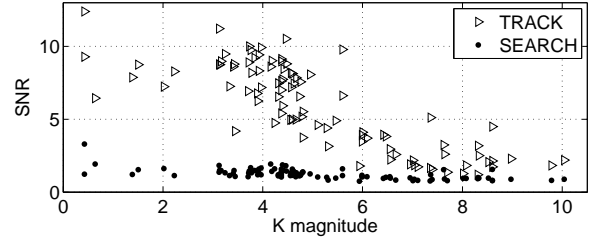} \end{center} 
\caption{Average SNR within the fringes (triangles, controller state TRACK) and in absence of fringes (dots, controller state SEARCH) for FSU observations (95 files). Error bars are suppressed for clarity.}
\label{fig:snr1}
\end{figure}

\begin{figure}
\begin{center} 
\includegraphics[width= \linewidth]{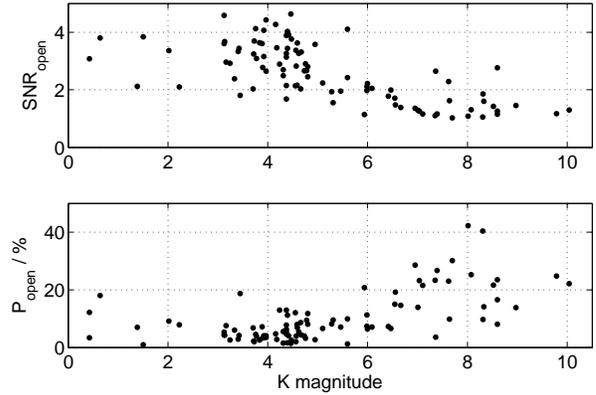} \end{center} 
\caption{SNR probability distribution intersection $SNR_\textit{open}$ (\textit{top}) and the associated probability $P_{open}$ (\textit{bottom}) as function of target magnitude (95 files).}
\label{fig:snr2}
\end{figure}

The FSU SNR as computed from Eq. \ref{eq:snr} is essential to allow the OPD controller to switch between states, and the quality of this estimator thus defines the capability of VLTI to track fringes on a given star. Figure \ref{fig:snr1} shows the SNR estimate as a function of target magnitude.\\ 
The SNR in Fig. \ref{fig:snr1} is computed as the mean of the SNR distributions $\Sigma_\mathrm{TRACK}$ and $\Sigma_\mathrm{SEARCH}$, each containing several thousand samples with and without fringes, respectively, obtained during the individual observation. We find that these distributions usually can be approximated Gaussian with standard deviations $\sigma_\mathrm{TRACK}$ and $\sigma_\mathrm{SEARCH}$. Let $SNR_\textit{open}$ be the value, where the probability distributions associated with $\Sigma_\mathrm{TRACK}$ and $\Sigma_\mathrm{SEARCH}$ have the same amplitude. In other words, $SNR_\textit{open}$ is the value where fringe detection and false-detection probabilities are equal to $P_\textit{open}$. Figure \ref{fig:snr2} shows these parameters. With increasing magnitude, $SNR_\textit{open}$ decreases, and the probability $P_\textit{open}$ rises, meaning that it becomes more difficult to distinguish fringes and noise based on the SNR estimate.\\
For our observations, the night calibration (Sect. \ref{sec:ncal}) was only partially operational and the SNR thresholds of the OPD controller (\textit{det}, \textit{close}, and \textit{open}, cf. Sect. \ref{sec:FSUftk}) had to be adjusted manually for each target, which is a time-consuming process that requires the experience of the observer. With the results presented in this section, we were able to define how to automatically set these thresholds.\\
The \textit{open} threshold is set equal to $SNR_\textit{open}$ to avoid confusing fringes with noise. \textit{det} is set to $\Sigma_\mathrm{TRACK} +  \sigma_\mathrm{TRACK}$, whereas \textit{close} is set to $\Sigma_\mathrm{TRACK} -  \sigma_\mathrm{TRACK}$. Since these parameters cannot be computed before fringes are detected, default values have to be defined. For $m_K<6$ these are $\textit{det} = 5$,  $\textit{close} = 3$, and $\textit{open} = 2$. For $m_K>6$ default values can be $\textit{det} = 3$,  $\textit{close} = 1.5$, and $\textit{open} = 1$, although the threshold setting becomes more difficult, because the probability of false detection increases as is visible in Fig. \ref{fig:snr2}.\\
The SNR bias due to the intensity level and the read-noise linked to the applied integration time is apparent in Fig. \ref{fig:snr1}: for short integration times and high intensities ($m_K<5$), $\Sigma_\mathrm{SEARCH}$ has an average value of $\sim$1.5, whereas it decreases with magnitude and reaches $\sim$0.8 for $m_K>7$. The effect is that the controller thresholds have to be adapted for the two regimes, as suggested above, which could be avoided by improving the calibration procedure and accounting for read- and photon noise in the SNR computation \citep{Colavita1999}. The estimates of visibility amplitude and SNR are related by a multiplicative factor and are therefore affected by the same biases (see Sect. \ref{sec:vis}).
\subsection{Dispersion and non-linearity effects}\label{sec:disp}
\begin{figure}
\begin{center} 
\includegraphics[width= \linewidth]{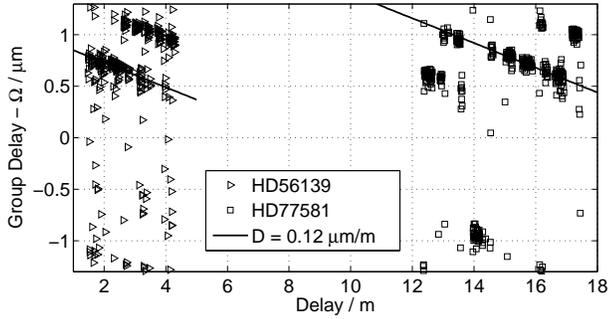} \end{center} 
\caption{Group delay minus FSU OPD as function of VLTI delay. Each point corresponds to 1~s of data.}
\label{fig:fringing}
\end{figure}

The difference between phase and group delay of a propagating wavepacket depends on the medium's dispersion. Since the FSU measures both phase and group delay, we can use it to examine air dispersion effects on interferometric observations, where we limit our discussion to the first-order, quasi-static effects. For VLTI observations, the vacuum delay between the coherent wavefronts arriving at the telescopes is compensated with delay lines in free air, which imposes dispersion of the wavepacket for non-zero delays. As the delay changes with earth rotation, the dispersion variation can be measured with the difference of phase and group delay \citep{Akeson:2000dq}. Figure \ref{fig:fringing} shows group delay minus phase as function of delay for two targets obtained during one night on a 96~m baseline. The total observing time per target is 15~min, which is too short to cover a variation of one fringe. However, we can measure the difference rate to $D=0.12\pm0.01 \, \mu\mathrm{m}/\mathrm{m}$, which is 5 times lower than the value measured with the Palomar testbed interferometer by \cite{Colavita2004}. For other nights, where $D$ can be estimated, we find values in the range of $D=0.10 \--0.14 \, \mu\mathrm{m}/\mathrm{m}$. As explained by \cite{Colavita2004}, the difference rate depends on temperature, pressure and humidity and is dominated in \textit{K}-band by water vapour dispersion to first-order. The relative humidity during the observations in Fig. \ref{fig:fringing} is 3~\% ($T=16.5^{\circ}$C and $P=744$ mbar), which may explain the low value.\\
Figure \ref{fig:fringing} also illustrates an effect that we note for a number of FSU observations: during fringe tracking, the average difference between phase and group delay\footnote{The controller is tracking on zero group delay, see Sect. \ref{sec:FSUftk}} not only changes due to dispersion but also exhibits several levels separated by $\sim0.5\mu$m. This is also apparent in Fig. \ref{fig:FTK} when carefully examining the FSU OPD values.\\
This is probably caused by a highly non-linear group delay estimate with a non-monotonic response function, which can be close to zero for several OPD positions. The controller can then either control to zero group delay, which is the more stable case, or lock on a meta-stable, non-zero position. The track-point typically changes after a fringe-loss event, when the controller returns to equilibrium. Since the phase and group delay response functions are different, the difference between phase and group delay depends on the track-point, which is reflected in the several levels in Fig. \ref{fig:fringing}.\\
Consequently, the FSU is not able to provide a unique track-point for the OPD controller, and this introduces biases in the measurements of the attached instrument. In principle, the SNR or visibility amplitude could be used to constrain the track-point error, but in practice the cyclic error (which for sky observations is larger than in Fig. \ref{fig:PDGD}) only allows us to conclude that the track-point is within the three central fringes.\\
The situation can be clarified and characterised when dual-feed operation becomes available: the phase and group delay linearity can then be measured on the target itself, and it is also possible to evaluate the FSU fringe-tracking performance independently and not, as throughout our analysis, based on the control signals themselves. In addition, the complete night calibration becomes available, which makes it possible to correct beam-train polarisation effects and consequently to reduce the cyclic errors and non-linearities.
\subsection{Fringe-tracking with UTs}
\begin{figure}
\begin{center} 
\includegraphics[width= \linewidth]{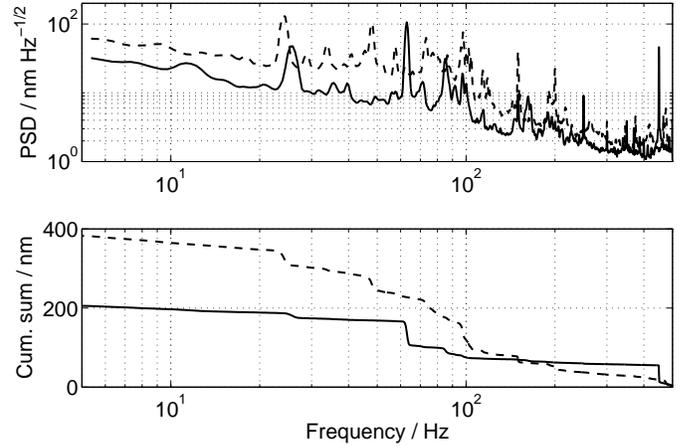} \end{center}
\caption{Comparison of OPD residual power spectral density above 5~Hz (\textit{top}) and its cumulative sum (\textit{bottom}) for observations with ATs (solid line) and UTs (dashed line) in normal conditions  ($\tau_0 = 3.4$~ms, AT observation run identical to Figs. \ref{fig:fluxpsd} and \ref{fig:opdpsd}). The cumulative sum equals the square root of the integral from the right over squared power spectral density.}
\label{fig:UTATpsd}
\end{figure}

Initial fringe tracking tests involving two 8~m VLT UTs were carried out over three hours during the night of 8 March 2009. Four stars were observed with UT3 and UT4 and fringes were observed with FSUA. The adaptive optics systems \citep{Arsenault:2004cr} and the vibration cancellation system based on accelerometers \citep{Haguenauer2008} were enabled. Fringe tracking was achieved for three stars with minimum brightness of $m_K = 8.6$, for the fourth star with apparent magnitude $m_K=11.7$ fringes were detected but not tracked (Table \ref{table:UT}).\\
As in the case of FINITO \citep{Haguenauer2008,Lieto2008}, FSU fringe tracking with the UTs is affected by structural vibrations of the telescope beamtrain. The OPD residuals of FSUA over one second are in the range of $300\--450$~nm RMS in the few cases presented, which is substantially higher than the corresponding numbers with the ATs. Figure \ref{fig:UTATpsd} compares the power spectral densities of FSU OPD for an observation with ATs and UTs. In the UT case the noise is constantly higher than in the AT case, except for few vibration peaks. The excess of low-frequency noise is caused by fringe-loss events, which occur frequently. Several strong features at 20\--100 Hz are visible and are known to originate in mechanical vibrations in the UTs \citep{Haguenauer2008}. Consequently the residual OPD jitter is higher, which considerably reduces the efficiency of observations with the UTs.\\  
Considering the very limited amount of collected data, all results presented in this section are preliminary. However, we demonstrated that the FSU is capable of fringe tracking with two UTs and faint objects can be reached. Further tests will be done in the near future and will then allow us to conclude on the operational limits.
\begin{table*} 
\caption{First FSU fringe-tracking targets with UTs}
\label{table:UT}  
\centering  
\begin{tabular}{c c l r c c c c c c}  
\hline\hline 
Date & Configuration & Target & $m_K$ & Baseline & Seeing & $\tau_0$ & DIT  & Lock ratio & residual\\
 &  &  & &  (m) &  ($\arcsec$) & (ms)  & (ms) &  (\%) & (nm RMS)\\ 
\hline  
8 Mar. 2009 & FSUA[1] & \object{HD94890} & 2.2 & 62 (UT3-UT4) & 0.6 & 6.3 & 1 &  100 & $\sim$ 330 \\
8 Mar. 2009 & FSUA[1] & \object{HD157591} & 4.5 & 62 (UT3-UT4) & 1.1 & 3.5 & 1 & 100 & $\sim$ 440\\
8 Mar. 2009 & FSUA[1] & \object{HD116714} & 8.6 & 62 (UT3-UT4) & 0.9 & 4.5 & 2 & 100 & $\sim$ 340\\
8 Mar. 2009 & FSUA[1] & \object{RX J1514.7-4220} & 11.7 & 62 (UT3-UT4) & 1.5 & 2.5 & 4 & $\sim$ 5 & $\sim$ 800 \\
\hline  
\end{tabular} 
\end{table*} 
\section{Discussion}
The described fringe tracking results do not constitute PRIMA observations. FSUA or FSUB is used as the only constituent of the PRIMA facility for on-axis fringe tracking with the VLTI infrastructure, and neither the star separator modules nor the differential delay lines are involved. The PRIMA metrology is solely used for the laboratory calibration.
\subsection{Operation range and performance}
In accordance with FINITO experience \citep{Bouquin2008}, fringe tracking with the FSU and the ATs is primarily limited by the injection efficiency. In bad atmospheric conditions, non-corrected tip-tilt perturbations cause frequent flux dropouts that prevent fringe tracking or severely degrade it. For the FSU there is an additional coupling between the tip-tilt and piston perturbations, caused by differential injection into the ABCD fibres and the absence of photometric channels. This coupling is transmitted by the piston feedback control loop. It is questionable that the real-time use of the total intensity sum for control can prevent this undesired effect, and it has instead to be seen as a feature of this system. The deterioration of e.g. the residual OPD caused by this coupling is difficult to quantify, but it does not prevent fringe tracking in nominal conditions.\\
The residual OPD during FSU fringe tracking shows a strong dependence on the visible coherence time, which makes it the primary parameter to characterise the atmospheric quality. The approximate lower limit during operation is 150~nm RMS and is achieved over a wide range of conditions, although lower values are measured in some cases. It is dominated by the atmospheric piston, attenuated according to the system's rejection function, by the coupling between residual tip-tilt and artificial phase specific to the FSU and by vibrations of beam-train elements. It is higher than the performance goal of 100~nm RMS, which reduces the anticipated observation efficiency.\\
The fringe-tracking operation range identified for the FSU in terms of atmospheric conditions is compatible with the one of FINITO ($\tau_0 > 2.5$~ms, seeing $<1.2\arcsec$, airmass $< 1.5$, \citealt{ESO2009}), which shows that both fringe trackers are similarly sensitive to these parameters. However, statistics for FINITO are far more extensive. Once more statistics have been built up for the FSU, so accurate comparison with FINITO becomes possible.\\ 
A major concern is the possible existence of several track-points, when fringe tracking with the FSU and the OPD controller. It can be clarified only by testing FSU fringe tracking within the PRIMA dual-feed facility, which makes it possible to calibrate the FSU on the target and to acquire independent delay measurements, hence to verify the FSU behaviour. If the problem is confirmed, modifications to the real-time algorithms and/or the calibration procedure are required.\\  
Coarse object visibilities can be derived from FSU observations, although the observed biases and limitations are not completely quantified. Precise visibility measurement capability is not part of the FSU specifications.\\
The FSU SNR estimates are used by the VLTI OPD controller to automate fringe-tracking observations. According to the range of observing conditions, the controller's default SNR thresholds need to be adapted. The quality of SNR estimates can be expected to improve by considering bias terms associated to photon and read-out noise, which are currently disregarded. An additional and possibly larger bias affecting visibility and SNR estimates originates in dynamic differential injection in the ABCD fibres. The capability of simultaneous phase and group delay measurement allows us to estimate the atmospheric dispersion due to water vapour, which is a limiting parameter for high-precision interferometric observations \citep{Colavita2004}.\\
Based on the current FSU magnitude limit for unresolved stars with the ATs of $m_K=9$ during commissioning, we can infer a realistic limit for routine operation to be $m_K \sim 7 \-- 8$, which is two magnitudes fainter than the current VLTI limit using FINITO \citep{ESO2009}. Some improvement in FSU performance can still be expected from software and operation changes, e.g. by taking the individual pixel wavelengths into account for phase and group delay computation, considering the different photometry of the input beams, and by implementing the on-object fringe calibration described in Sect. \ref{sec:fringe}. The OPD controller state machine can be optimised for operation with the FSU, as done for FINITO \citep{Bouquin2008}. In summary, the goal for PRIMA sensitivity of $m_K=8$ for the primary object \citep{Belle2008} appears to be within reach, although the absolute throughput of PRIMA is smaller than when only using the FSU, because of the additional reflections in the star separator modules and the differential delay lines.
\subsection{Limiting magnitude with ATs}
\cite{Quirrenbach:1998mi} describe the PRIMA conceptual study and derive the anticipated FSU fringe-tracking limiting magnitude of $m_K=12$ with ATs, based on unpublished assumptions. Without going into detail, we can evaluate the FSU performance with respect to these assumptions.\\
The total effective transmission of FSU and VLTI including wavefront errors is estimated from the observations to $4\pm1$~\% in \textit{K}-band, which is about one sixth of the assumed transmission of 25~\%. \cite{Quirrenbach:1998mi} assume a detector integration time of 20~ms and perfectly corrected, flat wavefronts coming from the telescopes. The current limit of $m_K=9$ is achieved with a shorter integration time of 4~ms, and we can interpolate the theoretical limit of $m_K=10.7$ for 20~ms, although fringe tracking for integration times above 10~ms is not demonstrated. The detector read-out noise is assumed to be 25 electrons RMS per pixel, which is comparable to the measured values of $31\pm4$ electrons RMS (per pixel) and $21\pm4$ electrons RMS (two-pixel difference).\\
In conclusion, the estimated FSU fringe-tracking sensitivity of $m_K=12$ with ATs is not reached. The discrepancy between predicted and measured limiting magnitude can be explained by the sensitivity loss from injection fluctuations caused by perturbed wavefronts entering the FSU and from the lower than anticipated effective transmission.
\section{Conclusions}
The FSU was integrated and aligned at the VLTI in July and August 2008. The successful and fast start of fringe tracking operation profited heavily from the preceding two-year period of intensive laboratory testing in Garching and the experience gained by our team during this time.\\
The FSU yields phase and group delay measurements at sampling rates up to 2 kHz, which are used to drive the fringe-tracking control loop. During the first three commissioning runs, it was used to track the fringes of stars with {\it K}-band magnitudes as faint as $m_K=9.0$, using two VLTI ATs and baselines of up to 96~m. Fringes on a fainter star of $m_K=10.0$ were recorded but not tracked. The FSU operation range is constrained with respect to atmospheric conditions. Further characterisation is required to conclude on the dependence of fringe-tracking performance on the target visibility. As of now, the VLTI incorporates two sensors for fringe tracking based on different detection principles. This is a unique opportunity to study and compare their performances, especially given the importance of fringe tracking for the upcoming second generation VLTI instruments.\\
Fringe tracking with two 8~m VLT UTs was demonstrated during one initial test run, and a target magnitude of $m_K = 8.6$ reached. Fringes on a star with $m_K=11.7$ were detected but not tracked, which constitutes the deepest observation with the VLTI to date. The concept of spatial phase modulation for fringe sensing and tracking in stellar interferometry is demonstrated for the first time with the FSU. During commissioning and combining light from two ATs, the FSU showed its ability to improve the VLTI fringe tracking sensitivity by more than one magnitude in {\it K}-band towards fainter objects, which is fundamental for achieving the scientific objectives of PRIMA.\\
However, many observation scenarios remain to be tested. The FSU behaviour within the PRIMA facility, more observations using the UTs and the response to future improvements of the VLTI infrastructure with respect to fringe-tracking bandwidth, vibration environment, and injection efficiency will reveal the ultimate FSU performance.
\begin{acknowledgements}
The authors acknowledge the major efforts of the PRIMA team and many staff members at ESO, that contributed to the development of the FSU. We thank B.~Delabre, F.~Derie, T.~Phan~Duc, P.~Duhoux, G.~Finger, R.~Frahm, S.~L{\'e}v\^eque, N.~Di~Lieto, L.~Mehrgan, J.-M.~Moresmau, N.~Schuhler, and A.~Wallander (now ITER).\\
Furthermore we are grateful to the Paranal Observatory staff, especially to the VLTI group and telescope operators, to G.~T.~van Belle and to L.~Jocou from LAOG for the collective support during the integration and commissioning of the FSU. P. Gitton, P. Haguenauer, and S. Morel provided invaluable help and initiative. We thank F.~Eisenhauer and his team at the Max Planck Institut f\"ur Extraterrestrische Physik for hosting the FSU testbed and for their helpful support. At Thales Alenia Space Italia and Osservatorio Astronomico di Torino, we would like to thank G.~Nicolini (now Alnilam) and G.~Massone. We thank G.~Vasisht for his assistance and for kindly providing the FSU group delay algorithm. For comments and contributions that improved the manuscript we thank C.~Schmid. J. S. thanks J.-B.~Le~Bouquin for stimulating discussions. We appreciate valuable comments made by the referee, which helped to improve the paper quality.\\
This research has made use of the Smithsonian/NASA Astrophysics Data System (ADS) and of the Centre de Donn{\'e}es astronomiques de Strasbourg (CDS).
\end{acknowledgements}
\bibliographystyle{aa} 
\bibliography{12271} 

\begin{thebibliography}{61}
\expandafter\ifx\csname natexlab\endcsname\relax\def\natexlab#1{#1}\fi

\bibitem[{{Abuter} {et~al.}(2008){Abuter}, {Popovic}, {Pozna}, {Sahlmann}, \&
  {Eisenhauer}}]{Abuter2008}
{Abuter}, R., {Popovic}, D., {Pozna}, E., {Sahlmann}, J., \& {Eisenhauer}, F.
  2008, in Proc. SPIE 7013

\bibitem[{{Abuter} {et~al.}(2006){Abuter}, {Rabien}, {Eisenhauer}, {Sahlmann},
  {Di Lieto}, {Haug}, {Wallander}, {L{\'e}v\^eque}, {M{\'e}nardi},
  {Delplancke}, {Schuhler}, {Kellner}, \& {Frahm}}]{Abuter2006}
{Abuter}, R., {Rabien}, S., {Eisenhauer}, F., {et~al.} 2006, in Proc. SPIE 6268

\bibitem[{{Akeson} {et~al.}(2000){Akeson}, {Swain}, \&
  {Colavita}}]{Akeson:2000dq}
{Akeson}, R.~L., {Swain}, M.~R., \& {Colavita}, M.~M. 2000, in Proc. SPIE 4006

\bibitem[{{Arsenault} {et~al.}(2004){Arsenault}, {Donaldson}, {Dupuy},
  {Fedrigo}, {Hubin}, {Ivanescu}, {Kasper}, {Oberti}, {Paufique}, {Rossi},
  {Silber}, {Delabre}, {Lizon}, \& {Gigan}}]{Arsenault:2004cr}
{Arsenault}, R., {Donaldson}, R., {Dupuy}, C., {et~al.} 2004, in Proc. SPIE
  5490

\bibitem[{{Barry} {et~al.}(2008){Barry}, {Danchi}, {Traub}, {Sokoloski},
  {Wisniewski}, {Serabyn}, {Kuchner}, {Akeson}, {Appleby}, {Bell}, {Booth},
  {Brandenburg}, {Colavita}, {Crawford}, {Creech-Eakman}, {Dahl}, {Felizardo},
  {Garcia}, {Gathright}, {Greenhouse}, {Herstein}, {Hovland}, {Hrynevych},
  {Koresko}, {Ligon}, {Mennesson}, {Millan-Gabet}, {Morrison}, {Palmer},
  {Panteleeva}, {Ragland}, {Shao}, {Smythe}, {Summers}, {Swain}, {Tsubota},
  {Tyau}, {Wetherell}, {Wizinowich}, {Woillez}, \& {Vasisht}}]{Barry2008}
{Barry}, R.~K., {Danchi}, W.~C., {Traub}, W.~A., {et~al.} 2008, \apj, 677, 1253

\bibitem[{{Bartko} {et~al.}(2008){Bartko}, {Pfuhl}, {Eisenhauer}, {Genzel},
  {Gillessen}, {Rabien}, {Abuter}, {v.~Belle}, {Delplancke}, {Menardi}, \&
  {Sahlmann}}]{Bartko2008}
{Bartko}, H., {Pfuhl}, O., {Eisenhauer}, F., {et~al.} 2008, in Proc. SPIE 7013

\bibitem[{{Bean} {et~al.}(2007){Bean}, {McArthur}, {Benedict}, {Harrison},
  {Bizyaev}, {Nelan}, \& {Smith}}]{Bean2007}
{Bean}, J.~L., {McArthur}, B.~E., {Benedict}, G.~F., {et~al.} 2007, \aj, 134,
  749

\bibitem[{{Benedict} {et~al.}(2002){Benedict}, {McArthur}, {Forveille},
  {Delfosse}, {Nelan}, {Butler}, {Spiesman}, {Marcy}, {Goldman}, {Perrier},
  {Jefferys}, \& {Mayor}}]{Benedict:2002rz}
{Benedict}, G.~F., {McArthur}, B.~E., {Forveille}, T., {et~al.} 2002, \apjl,
  581, L115

\bibitem[{{Berger} {et~al.}(2008){Berger}, {Monnier}, {Millan-Gabet}, {ten
  Brummelaar}, {Anderson}, {Blum}, {Blasius}, {Pedretti}, \&
  {Thureau}}]{Berger2008}
{Berger}, D.~H., {Monnier}, J.~D., {Millan-Gabet}, R., {et~al.} 2008, in Proc.
  SPIE 7013

\bibitem[{{Bonnet} {et~al.}(2006){Bonnet}, {Bauvir}, {Wallander}, {Cantzler},
  {Carstens}, {Caruso}, {di Lieto}, {Guisard}, {Haguenauer}, {Housen},
  {Mornhinweg}, {Nicoud}, {Ramirez}, {Sahlmann}, {Vasisht}, {Wehner}, \&
  {Zagal}}]{Bonnet2006}
{Bonnet}, H., {Bauvir}, B., {Wallander}, A., {et~al.} 2006, The Messenger, 126,
  37

\bibitem[{{Colavita}(1994)}]{Colavita1994}
{Colavita}, M.~M. 1994, \aap, 283, 1027

\bibitem[{{Colavita}(1999)}]{Colavita1999}
{Colavita}, M.~M. 1999, \pasp, 111, 111

\bibitem[{{Colavita} {et~al.}(2004){Colavita}, {Swain}, {Akeson}, {Koresko}, \&
  {Hill}}]{Colavita2004}
{Colavita}, M.~M., {Swain}, M.~R., {Akeson}, R.~L., {Koresko}, C.~D., \&
  {Hill}, R.~J. 2004, \pasp, 116, 876

\bibitem[{{Colavita} {et~al.}(1999){Colavita}, {Wallace}, {Hines}, {Gursel},
  {Malbet}, {Palmer}, {Pan}, {Shao}, {Yu}, {Boden}, {Dumont}, {Gubler},
  {Koresko}, {Kulkarni}, {Lane}, {Mobley}, \& {van Belle}}]{Colavita1999a}
{Colavita}, M.~M., {Wallace}, J.~K., {Hines}, B.~E., {et~al.} 1999, \apj, 510

\bibitem[{{Creath}(1988)}]{Creath1988}
{Creath}, K. 1988, in {Progress in Optics}, ed. E.~{Wolf}, Vol. XXVI, 349--393

\bibitem[{{Daigne} \& {Lestrade}(1999)}]{Daigne:1999sf}
{Daigne}, G. \& {Lestrade}, J.-F. 1999, \aaps, 138, 355

\bibitem[{{Delplancke} {et~al.}(2006){Delplancke}, {Derie}, {L{\'e}v\^eque},
  {M{\'e}nardi}, {Abuter}, {Andolfato}, {Ballester}, {de Jong}, {Di Lieto},
  {Duhoux}, {Frahm}, {Gitton}, {Glindemann}, {Palsa}, {Puech}, {Sahlmann},
  {Schuhler}, {Duc}, {Valat}, \& {Wallander}}]{Delplancke2006}
{Delplancke}, F., {Derie}, F., {L{\'e}v\^eque}, S., {et~al.} 2006, in Proc.
  SPIE 6268

\bibitem[{{Di Lieto} {et~al.}(2008){Di Lieto}, {Haguenauer}, {Sahlmann}, \&
  {Vasisht}}]{Lieto2008}
{Di Lieto}, N., {Haguenauer}, P., {Sahlmann}, J., \& {Vasisht}, G. 2008, in
  Proc. SPIE 7013

\bibitem[{{Elias} {et~al.}(2008){Elias}, {K{\"o}hler}, {Stilz}, {Reffert},
  {Geisler}, {Quirrenbach}, {de Jong}, {Delplancke}, {Tubbs}, {Launhardt},
  {Henning}, {M{\'e}gevand}, \& {Queloz}}]{Elias:2008sf}
{Elias}, II, N.~M., {K{\"o}hler}, R., {Stilz}, I., {et~al.} 2008, in Proc. SPIE
  7013

\bibitem[{ESO(2009)}]{ESO2009}
ESO. 2009, {Call for proposals - P84}

\bibitem[{{Ferrari} {et~al.}(2003){Ferrari}, {Lemaitre}, {Mazzanti}, {Derie},
  {Huxley}, {Lemerrer}, {Lanzoni}, {Dargent}, \& {Wallander}}]{Ferrari:2003qe}
{Ferrari}, M., {Lemaitre}, G.~R., {Mazzanti}, S.~P., {et~al.} 2003, in Proc.
  SPIE 4838

\bibitem[{{Gai} {et~al.}(2004){Gai}, {Menardi}, {Cesare}, {Bauvir}, {Bonino},
  {Corcione}, {Dimmler}, {Massone}, {Reynaud}, \& {Wallander}}]{Gai2004}
{Gai}, M., {Menardi}, S., {Cesare}, S., {et~al.} 2004, in Proc. SPIE 5491

\bibitem[{{Gitton} \& {Puech}(2009)}]{gitton2009}
{Gitton}, P. \& {Puech}, F. 2009, Interface Control Document between VLTI and
  its Instruments, Issue 6.0, Tech. rep., ESO

\bibitem[{{Gitton} {et~al.}(2004){Gitton}, {Leveque}, {Avila}, \& {Phan
  Duc}}]{Gitton:2004wd}
{Gitton}, P.~B., {Leveque}, S.~A., {Avila}, G., \& {Phan Duc}, T. 2004, in
  Proc. SPIE 5491

\bibitem[{{Haguenauer} {et~al.}(2008){Haguenauer}, {Abuter}, {Alonso},
  {Argomedo}, {Bauvir}, {Blanchard}, {Bonnet}, {Brillant}, {Cantzler}, {Derie},
  {Delplancke}, {Di Lieto}, {Dupuy}, {Durand}, {Gitton}, {Gilli}, {Glindemann},
  {Guniat}, {Guisard}, {Haddad}, {Hudepohl}, {Hummel}, {Jesuran}, {Kaufer},
  {Koehler}, {Le Bouquin}, {L{\'e}v\^eque}, {Lidman}, {Mardones},
  {M{\'e}nardi}, {Morel}, {Percheron}, {Petr-Gotzens}, {Phan Duc}, {Puech},
  {Ramirez}, {Rantakyr{\"o}}, {Richichi}, {Rivinius}, {Sahlmann}, {Sandrock},
  {Sch{\"o}ller}, {Schuhler}, {Somboli}, {Stefl}, {Tapia}, {Van Belle},
  {Wallander}, {Wehner}, \& {Wittkowski}}]{Haguenauer2008}
{Haguenauer}, P., {Abuter}, R., {Alonso}, J., {et~al.} 2008, in Proc. SPIE 7013

\bibitem[{{Koehler} {et~al.}(2006){Koehler}, {Kraus}, {Moresmau},
  {Wirenstrand}, {Duhoux}, {Karban}, {Andolfato}, \& {Gonte}}]{Koehler:2006rr}
{Koehler}, B., {Kraus}, M., {Moresmau}, J.~M., {et~al.} 2006, in Proc. SPIE
  6268

\bibitem[{{Lane} \& {Colavita}(2003)}]{Lane:2003zl}
{Lane}, B.~F. \& {Colavita}, M.~M. 2003, \aj, 125, 1623

\bibitem[{{Lane} \& {Muterspaugh}(2004{\natexlab{a}})}]{Lane:2004rm}
{Lane}, B.~F. \& {Muterspaugh}, M.~W. 2004{\natexlab{a}}, \apj, 601, 1129

\bibitem[{{Lane} \& {Muterspaugh}(2004{\natexlab{b}})}]{Lane:2004rz}
{Lane}, B.~F. \& {Muterspaugh}, M.~W. 2004{\natexlab{b}}, in Proc. SPIE 5491

\bibitem[{{Lane} {et~al.}(2007){Lane}, {Muterspaugh}, {Fekel}, {Williamson},
  {Browne}, {Konacki}, {Burke}, {Colavita}, {Kulkarni}, \&
  {Shao}}]{Lane:2007sf}
{Lane}, B.~F., {Muterspaugh}, M.~W., {Fekel}, F.~C., {et~al.} 2007, \apj, 669,
  1209

\bibitem[{{Launhardt} {et~al.}(2008){Launhardt}, {Queloz}, {Henning},
  {Quirrenbach}, {Delplancke}, {Andolfato}, {Baumeister}, {Bizenberger},
  {Bleuler}, {Chazelas}, {D{\'e}rie}, {Di Lieto}, {Duc}, {Duvanel}, {Elias},
  {Fluery}, {Geisler}, {Gillet}, {Graser}, {Koch}, {K{\"o}hler}, {Maire},
  {M{\'e}gevand}, {Michellod}, {Moresmau}, {M{\"u}ller}, {M{\"u}llhaupt},
  {Naranjo}, {Pepe}, {Reffert}, {Sache}, {S{\'e}gransan}, {Salvad{\'e}},
  {Schulze-Hartung}, {Setiawan}, {Simond}, {Sosnowska}, {Stilz}, {Tubbs},
  {Wagner}, {Weber}, {Weise}, \& {Zago}}]{Launhardt2008}
{Launhardt}, R., {Queloz}, D., {Henning}, T., {et~al.} 2008, in Proc. SPIE 7013

\bibitem[{{Lawson} {et~al.}(2000){Lawson}, {Colavita}, {Dumont}, \&
  {Lane}}]{Lawson:2000sf}
{Lawson}, P.~R., {Colavita}, M.~M., {Dumont}, P.~J., \& {Lane}, B.~F. 2000, in
  Proc. SPIE 4006

\bibitem[{{Le Bouquin} {et~al.}(2008){Le Bouquin}, {Abuter}, {Bauvir},
  {Bonnet}, {Haguenauer}, {di Lieto}, {Menardi}, {Morel}, {Rantakyr{\"o}},
  {Schoeller}, {Wallander}, \& {Wehner}}]{Bouquin2008}
{Le Bouquin}, J.-B., {Abuter}, R., {Bauvir}, B., {et~al.} 2008, in Proc. SPIE
  7013

\bibitem[{{Le Bouquin} {et~al.}(2009{\natexlab{a}}){Le Bouquin}, {Abuter},
  {Haguenauer}, {Bauvir}, {Popovic}, \& {Pozna}}]{Bouquin2009b}
{Le Bouquin}, J.-B., {Abuter}, R., {Haguenauer}, P., {et~al.}
  2009{\natexlab{a}}, \aap, 493, 747

\bibitem[{{Le Bouquin} {et~al.}(2009{\natexlab{b}}){Le Bouquin}, {Lacour},
  {Renard}, {Thi{\'e}baut}, {Merand}, \& {Verhoelst}}]{Bouquin2009a}
{Le Bouquin}, J.-B., {Lacour}, S., {Renard}, S., {et~al.} 2009{\natexlab{b}},
  \aap, 496, L1

\bibitem[{{Leinert} {et~al.}(2003){Leinert}, {Graser}, {Richichi},
  {Sch{\"o}ller}, {Waters}, {Perrin}, {Jaffe}, {Lopez}, {Glazenborg-Kluttig},
  {Przygodda}, {Morel}, {Biereichel}, {Haddad}, {Housen}, \&
  {Wallander}}]{Leinert2003}
{Leinert}, C., {Graser}, U., {Richichi}, A., {et~al.} 2003, The Messenger, 112,
  13

\bibitem[{{L{\'e}v{\^e}que} {et~al.}(1996){L{\'e}v{\^e}que}, {Koehler}, \& {von
  der L{\"u}he}}]{Leveque:1996rz}
{L{\'e}v{\^e}que}, S., {Koehler}, B., \& {von der L{\"u}he}, O. 1996, \apss,
  239, 305

\bibitem[{{Lindegren}(1980)}]{Lindegren:1980bu}
{Lindegren}, L. 1980, \aap, 89, 41

\bibitem[{{Linfield} {et~al.}(2001){Linfield}, {Colavita}, \&
  {Lane}}]{Linfield:2001rr}
{Linfield}, R.~P., {Colavita}, M.~M., \& {Lane}, B.~F. 2001, \apj, 554, 505

\bibitem[{{Meyer} {et~al.}(1998){Meyer}, {Finger}, {Mehrgan}, {Nicolini}, \&
  {Stegmeier}}]{Meyer1998}
{Meyer}, M., {Finger}, G., {Mehrgan}, H., {Nicolini}, G., \& {Stegmeier}, J.
  1998, in Proc. SPIE 3354

\bibitem[{{Monnier}(2003)}]{Monnier2003}
{Monnier}, J.~D. 2003, Reports on Progress in Physics, 66, 789

\bibitem[{{Morel} {et~al.}(2004){Morel}, {Vannier}, {Menardi},
  {Biancat-Marchet}, {Fischer}, {Gitton}, {Glindemann}, {Guisard}, {Haddad},
  {Housen}, {Huxley}, {Kiekebusch}, {Longinotti}, {Phan Duc}, {Schoeller}, \&
  {Wallander}}]{Morel:2004si}
{Morel}, S., {Vannier}, M., {Menardi}, S., {et~al.} 2004, in Proc. SPIE 5491

\bibitem[{{Mottini} {et~al.}(2005){Mottini}, {Cesare}, \&
  {Nicolini}}]{Mottini2005}
{Mottini}, S., {Cesare}, S., \& {Nicolini}, G. 2005, in Proc. SPIE 5962

\bibitem[{{Muterspaugh} {et~al.}(2008){Muterspaugh}, {Lane}, {Fekel},
  {Konacki}, {Burke}, {Kulkarni}, {Colavita}, {Shao}, \&
  {Wiktorowicz}}]{Muterspaugh:2008yg}
{Muterspaugh}, M.~W., {Lane}, B.~F., {Fekel}, F.~C., {et~al.} 2008, \aj, 135,
  766

\bibitem[{{Muterspaugh} {et~al.}(2005){Muterspaugh}, {Lane}, {Konacki},
  {Burke}, {Colavita}, {Kulkarni}, \& {Shao}}]{Muterspaugh:2005lq}
{Muterspaugh}, M.~W., {Lane}, B.~F., {Konacki}, M., {et~al.} 2005, \aj, 130,
  2866

\bibitem[{{Nijenhuis} {et~al.}(2008){Nijenhuis}, {Visser}, {de Man}, {Dekker},
  {Mekking}, \& {Kamphues}}]{Nijenhuis:2008cy}
{Nijenhuis}, J., {Visser}, H., {de Man}, H., {et~al.} 2008, in Proc. SPIE 7013

\bibitem[{{Pepe} {et~al.}(2008){Pepe}, {Queloz}, {Henning}, {Quirrenbach},
  {Delplancke}, {Andolfato}, {Baumeister}, {Bizenberger}, {Bleuler},
  {Chazelas}, {D{\'e}rie}, {Di Lieto}, {Duc}, {Duvanel}, {Fleury}, {Gillet},
  {Graser}, {Koch}, {Launhardt}, {Maire}, {M{\'e}gevand}, {Michellod},
  {Moresmau}, {M{\"u}llhaupt}, {Naranjo}, {Sache}, {Salvad{\'e}}, {Simond},
  {Sosnowska}, {Wagner}, \& {Zago}}]{Pepe2008}
{Pepe}, F., {Queloz}, D., {Henning}, T., {et~al.} 2008, in Proc. SPIE 7013

\bibitem[{{Petrov} {et~al.}(2007){Petrov}, {Malbet}, {Weigelt}, {Antonelli},
  {Beckmann}, {Bresson}, {Chelli}, {Dugu{\'e}}, {Duvert}, {Gennari},
  {Gl{\"u}ck}, {Kern}, {Lagarde}, {Le Coarer}, {Lisi}, {Millour}, {Perraut},
  {Puget}, {Rantakyr{\"o}}, {Robbe-Dubois}, {Roussel}, {Salinari}, {Tatulli},
  {Zins}, {Accardo}, {Acke}, {Agabi}, {Altariba}, {Arezki}, {Aristidi},
  {Baffa}, {Behrend}, {Bl{\"o}cker}, {Bonhomme}, {Busoni}, {Cassaing},
  {Clausse}, {Colin}, {Connot}, {Delboulb{\'e}}, {Domiciano de Souza},
  {Driebe}, {Feautrier}, {Ferruzzi}, {Forveille}, {Fossat}, {Foy},
  {Fraix-Burnet}, {Gallardo}, {Giani}, {Gil}, {Glentzlin}, {Heiden},
  {Heininger}, {Hernandez Utrera}, {Hofmann}, {Kamm}, {Kiekebusch}, {Kraus},
  {Le Contel}, {Le Contel}, {Lesourd}, {Lopez}, {Lopez}, {Magnard}, {Marconi},
  {Mars}, {Martinot-Lagarde}, {Mathias}, {M{\`e}ge}, {Monin}, {Mouillet},
  {Mourard}, {Nussbaum}, {Ohnaka}, {Pacheco}, {Perrier}, {Rabbia}, {Rebattu},
  {Reynaud}, {Richichi}, {Robini}, {Sacchettini}, {Schertl}, {Sch{\"o}ller},
  {Solscheid}, {Spang}, {Stee}, {Stefanini}, {Tallon}, {Tallon-Bosc}, {Tasso},
  {Testi}, {Vakili}, {von der L{\"u}he}, {Valtier}, {Vannier}, \&
  {Ventura}}]{Petrov2007}
{Petrov}, R.~G., {Malbet}, F., {Weigelt}, G., {et~al.} 2007, \aap, 464, 1

\bibitem[{{Pravdo} \& {Shaklan}(2009)}]{Pravdo:2009rz}
{Pravdo}, S.~H. \& {Shaklan}, S.~B. 2009, \apj, 700, 623

\bibitem[{{Quirrenbach} {et~al.}(1998){Quirrenbach}, {Coude Du Foresto},
  {Daigne}, {Hofmann}, {Hofmann}, {Lattanzi}, {Osterbart}, {Le Poole},
  {Queloz}, \& {Vakili}}]{Quirrenbach:1998mi}
{Quirrenbach}, A., {Coude Du Foresto}, V., {Daigne}, G., {et~al.} 1998, in
  Proc. SPIE 3350

\bibitem[{{Sahlmann}(2007)}]{Sahlmann2007a}
{Sahlmann}, J. 2007, Master's thesis, Albert-Ludwigs-Universit\"at, Freiburg im
  Breisgau, 2007.

\bibitem[{{Sahlmann} {et~al.}(2008{\natexlab{a}}){Sahlmann}, {Abuter}, {Di
  Lieto}, {M{\'e}nardi}, {Delplancke}, {Bartko}, {Eisenhauer}, {L{\'e}v\^eque},
  {Pfuhl}, {Schuhler}, {van Belle}, \& {Vasisht}}]{Sahlmann2008a}
{Sahlmann}, J., {Abuter}, R., {Di Lieto}, N., {et~al.} 2008{\natexlab{a}}, in
  Proc. SPIE 7013

\bibitem[{{Sahlmann} {et~al.}(2008{\natexlab{b}}){Sahlmann}, {Abuter},
  {M{\'e}nardi}, \& {Vasisht}}]{Sahlmann2008b}
{Sahlmann}, J., {Abuter}, R., {M{\'e}nardi}, S., \& {Vasisht}, G.
  2008{\natexlab{b}}, in IAU Symposium, Vol. 248, 124--125

\bibitem[{{Schuhler}(2007)}]{Schuhler2007}
{Schuhler}, N. 2007, PhD thesis, Universit{\'e} de Strasbourg

\bibitem[{{Shao} \& {Colavita}(1992)}]{Shao1992}
{Shao}, M. \& {Colavita}, M.~M. 1992, \aap, 262, 353

\bibitem[{{Shao} \& {Staelin}(1977)}]{Shao1977}
{Shao}, M. \& {Staelin}, D.~H. 1977, Journal of the Optical Society of America
  (1917-1983), 67, 81

\bibitem[{{Shao} \& {Staelin}(1980)}]{Shao1980}
{Shao}, M. \& {Staelin}, D.~H. 1980, \ao, 19, 1519

\bibitem[{{Tango}(1990)}]{Tango:1990rm}
{Tango}, W.~J. 1990, \ao, 29, 516

\bibitem[{{van Belle} {et~al.}(2008){van Belle}, {Sahlmann}, {Abuter},
  {Accardo}, {Andolfato}, {Brillant}, {de Jong}, {Derie}, {Delplancke}, {Duc},
  {Dupuy}, {Gilli}, {Gitton}, {Haguenauer}, {Jocou}, {Jost}, {di Lieto},
  {Frahm}, {M{\'e}nardi}, {Morel}, {Moresmau}, {Palsa}, {Popovic}, {Pozna},
  {Puech}, {L{\'e}v{\^e}que}, {Ramirez}, {Schuhler}, {Somboli}, {Wehner}, \&
  {The Espri Consortium}}]{Belle2008}
{van Belle}, G.~T., {Sahlmann}, J., {Abuter}, R., {et~al.} 2008, The Messenger,
  134, 6

\bibitem[{{Vasisht} {et~al.}(2003){Vasisht}, {Booth}, {Colavita}, {Johnson},
  {Ligon}, {Moore}, \& {Palmer}}]{Vasisht2003}
{Vasisht}, G., {Booth}, A.~J., {Colavita}, M.~M., {et~al.} 2003, in Proc. SPIE
  4838

\bibitem[{{Wyant}(1975)}]{Wyant1975}
{Wyant}, J.~C. 1975, \ao, 14, 2622

\end{thebibliography}
\end{document}